\documentclass[12pt,preprint]{aastex}
\usepackage{graphicx}
\usepackage{amsmath}
\usepackage{url}

\defcitealias{assef08}{A08}
\defcitealias{assef09}{Paper~II}

\begin{document}

\title{Low Resolution Spectral Templates For AGNs and Galaxies From
0.03 -- 30$\mu$m}

\author{R.J.~Assef\altaffilmark{1}, 
  C.S.~Kochanek\altaffilmark{1},
  M.~Brodwin\altaffilmark{2,3},
  R.~Cool\altaffilmark{4},
  W.~Forman\altaffilmark{2},
  A.H.~Gonzalez\altaffilmark{5},
  R.C.~Hickox\altaffilmark{2},
  C.~Jones\altaffilmark{2},
  E.~Le~Floc'h\altaffilmark{6},
  J.~Moustakas\altaffilmark{7},
  S.S.~Murray\altaffilmark{2}
  D.~Stern\altaffilmark{8}
}

\affil{
  \altaffiltext{1} {Department of Astronomy, The Ohio State
    University, 140 W.\ 18th Ave., Columbus, OH 43210
    [email:{\tt{rjassef@astronomy.ohio-state.edu}}]}
  \altaffiltext{2} {Harvard-Smithsonian Center for Astrophysics, 60
    Garden St., Cambridge, MA 02138}
  \altaffiltext{3} {W. M. Keck Postdoctoral Fellow at the
    Harvard-Smithsonian Center for Astrophysics}
  \altaffiltext{4} {Peyton Hall, Princeton University, Princeton, NJ
    08540}
  \altaffiltext{5} {Department of Astronomy, Bryant Space Science
    Center, University of Florida, Gainesville, FL 32611}
  \altaffiltext{6} {CEA-Saclay, Service d'Astrophysique, Orme des
    Merisiers, Bat.709, 91191 Gif-sur-Yvette, FRANCE}
  \altaffiltext{7} {Center for Astrophysics and Space Sciences
    University of California, San Diego 9500 Gilman Drive La Jolla,
    California, 92093-0424}
  \altaffiltext{8} {Jet Propulsion Laboratory, California Institute of
    Technology, 4800 Oak Grove Drive, Mail Stop 169-506, Pasadena, CA
    91109}
}

\begin{abstract}

We present a set of low resolution empirical SED templates for AGNs
and galaxies in the wavelength range from 0.03 to 30$\mu$m based on
the multi-wavelength photometric observations of the NOAO Deep-Wide
Field Survey Bo\"otes field and the spectroscopic observations of the
AGN and Galaxy Evolution Survey. Our training sample is comprised of
14448 galaxies in the redshift range $0 \lesssim z \lesssim 1$ and
5347 likely AGNs in the range $0\lesssim z \lesssim 5.58$. The galaxy
templates correspond to the SED templates presented by \citet{assef08}
extended into the UV and mid-IR by the addition of FUV and NUV GALEX
and MIPS 24$\mu$m data for the field. We use our templates to
determine photometric redshifts for galaxies and AGNs. While they are
relatively accurate for galaxies ($\sigma_z/(1+z) = 0.04$, with 5\%
outlier rejection), their accuracies for AGNs are a strong function of
the luminosity ratio between the AGN and galaxy components. Somewhat
surprisingly, the relative luminosities of the AGN and its host are
well determined even when the photometric redshift is significantly in
error. We also use our templates to study the mid-IR AGN selection
criteria developed by \citet{stern05} and \citet{lacy04}. We find that
the \citet{stern05} criteria suffers from significant incompleteness
when there is a strong host galaxy component and at $z\simeq 4.5$,
when the broad H$\alpha$ emission line is redshifted into the [3.6]
band, but that it is little contaminated by low and intermediate
redshift galaxies. The \citet{lacy04} criterion is not affected by
incompleteness at $z\simeq 4.5$ and is somewhat less affected by
strong galaxy host components, but is heavily contaminated by low
redshift star forming galaxies. Finally, we use our templates to
predict the color-color distribution of sources in the upcoming WISE
mission and define a color criterion to select AGNs analogous to those
developed for IRAC photometry. We estimate that in between $640,000$
and $1,700,000$ AGNs will be identified by these criteria, but without
additional information, WISE-selected quasars will have serious
completeness problems for $z\gtrsim 3.4$.
\end{abstract}

\keywords{galaxies: active --- galaxies: distances and redshifts ---
galaxies: photometry --- quasars: general}

\section{Introduction}\label{sec:intro}

Upcoming large photometric surveys like the ground based LSST
\citep{tyson02}, Pan-STARRS \citep{kaiser04}, DES \citep{des} and
UKIDSS \citep{lawrence07} projects, or the space-based WISE
\citep{mainzer05} mission, will increase the number of cataloged
sources in the universe by a large factor. Most objects always lie at
the survey magnitude limit, which means that spectroscopic follow-up
will only be possible for a small fraction of objects, and much of the
science will need to rely solely on multi-wavelength photometric
observations. One of the main tools for analyzing the multi-wavelength
photometry is the modeling of the spectral energy distributions (SED)
of the sources. SEDs can be used to estimate photometric redshifts and
K-corrections, to study luminosity functions and the properties of
different populations of objects, and to select smaller subsamples for
follow-up studies.

Galaxy SEDs have been studied using many different
techniques. Theoretical models based on population synthesis of
stellar spectra \citep[e.g.][]{bc03,pegase,starburst99} have been
successfully applied in many studies of galaxies. Their strength is
that they allow the estimation of physically relevant parameters, like
star formation rates, ages, masses and metallicities solely by
comparing the SED models to the photometric observations. However,
there are unknowns in the spectral properties of stars and stellar
populations, such as the presence of thermally pulsating AGB stars
\citep[e.g.][]{conroy09} and the effects of binaries
\citep{eldridge09}, as well as non-stellar sources of emission, such
as dust in the interstellar medium (ISM) or active nuclei, and
absorption, both broadly distributed and associated with stars, that
alter or modify the SEDs. Some of these issues can be avoided by the
use of empirical SED templates
\citep[e.g.][]{cww,kinney96,devriendt99,assef08}. However, most
empirical templates lack the broad wavelength coverage of the
theoretical ones.

Almost all AGN SED templates are empirical, as the physical processes
governing AGN emission are not quantitatively
understood. \citet{elvis94} constructed the mean SED of Type 1 AGNs by
combining spectroscopic redshifts and multi-wavelength photometric
observations ranging from the radio to the X-rays of 47 known
quasars. A similar approach was followed by \citet{richards06a} over a
similar wavelength range based on 259 SDSS Type 1 AGNs. Both studies,
however, include a limited number of AGNs, do not consider reddening
and do not self-consistently model contamination from the host galaxy.

In a previous paper \citep[][hereafter \citetalias{assef08}]{assef08}
we presented a set of empirical galaxy templates derived from the
extensive photometric observations of the NOAO Deep Wide-Field Survey
\citep[NDWFS;][]{ndwfs99} Bo\"otes field and their spectroscopic
follow up observations by the AGN and Galaxy Evolution Survey
\citep[AGES;][]{kochanek09}. These templates form a linear
non-negative basis for the color-color space spanned by galaxies and
accurately reproduce SEDs in the wavelength range from 0.2 to
10$\mu$m. Here, we extend that work in two different
directions. First, we extend the wavelength range from 0.03 to 30 $\mu$m
by incorporating GALEX UV \citep{martin05} and MIPS 24$\mu$m
\citep{weedman06} observations, and secondly, we add a template for
AGN emission derived in the same manner as the galaxy templates. By
simultaneously deriving the AGN and galaxy SEDs, we self-consistently
account for the host contamination. We also correct for the reddening
of the different AGNs in our sample. Because AGES selects AGN using a
broad range of selection methods, our AGN template is not exclusively
applicable to unreddened Type 1 quasars. Earlier versions of this
extended set of templates have already been used by \citet{atlee09} as
a part of a study of the evolution of the UV upturn in elliptical
galaxies, and by \citet{ross09} to study the resolved host of the
$z=1.7$ gravitationally lensed quasar SDSS J1004+4112. The latter
study helped to verify that our models could correctly separate host
and AGN contributions for unresolved systems.

This paper is structured as follows. In \S\ref{sec:data} we describe
our photometric and spectroscopic data set. The algorithms we use to
obtain the SED templates are discussed in \S\ref{sec:methods}, as well
as the algorithm to use them for estimating photometric redshifts. In
\S\ref{sec:results} we discuss our SED templates and their accuracy
for estimating photometric redshifts of galaxies and AGNs, and we
apply them to study the problems of mid-IR QSO selection in the IRAC
bands and in the upcoming WISE satellite mission. In a subsequent
paper \citep[][\citetalias{assef09}]{assef09} we use these templates
to study the IRAC-selected QSO luminosity function and its evolution
with redshift. Throughout the text we assume a $\Lambda$CDM cosmology
with $H_0 = 73~\rm km~\rm s^{-1}~\rm Mpc^{-1}$, $\Omega_{\rm M} =
0.3$ and $\Omega_{\Lambda} = 0.7$.

\section{Data}\label{sec:data}

The data used in this work are an expansion over that of
\citetalias{assef08}. It is based on the extensive multi-wavelength
imaging of the NOAO Deep Wide-Field Survey \citep[NDWFS;][]{ndwfs99}
Bo\"otes field and the spectroscopic observations of the same field by
the AGN and Galaxy Evolution Survey \citep[AGES;][]{kochanek09}. NDWFS
is a deep optical and near-infrared survey that covers two 9.3 deg$^2$
regions of the sky, the Bo\"otes and Cetus fields. We focus on the
Bo\"otes field, which was imaged in $B_{W}$ (3500-4750 \AA, peak at
$\approx 4000$ \AA), $R$, $I$ and $K$ pass-bands to depths of
approximately 26.5, 26, 25.5 and 21 AB magnitudes ($5 \sigma$, 2$''$
diameter aperture). Extensive follow-up imaging of the field at
different wavelengths has been obtained by several other surveys. In
particular we include $J$ and $K_s$ data from the Flamingos
Extragalactic Survey \citep[FLAMEX;][]{flamex06}, $z'$ data from
zBo\"otes \citep{cool06}, MIPS 24$\mu$m data from \citet{weedman06},
NUV and FUV observations from the GALEX GR5 release
\citep{morrissey07}, and the mid-IR IRAC observations of the Spitzer
Deep-Wide Field Survey \citep[SDWFS;][]{ashby09}. The IRAC Shallow
Survey \citep{eisenhardt04}, which we used in \citetalias{assef08}, is
a subset of the SDWFS data. While in \citetalias{assef08} we referred
to the IRAC channels by C1, C2, C3 and C4 for the 3.6, 4.5, 5.8, and
8$\mu$m bands, here we will use the more common nomenclature with the
channel's wavelength in square brackets ([3.6],[4.5],[5.8] and
[8.0]). It should be noted that there are also radio (FIRST:
\citealt{first}; WENSS: \citealt{wenss}; WSRT: \citealt{wsrt}; NVSS:
\citealt{nvss}) and X-ray ({\it{Chandra}} XBo\"otes:
\citealt{xbootes}) observations of the field that we do not directly
include in our analysis.

AGES is a redshift survey in the NDWFS Bo\"otes field using the 300
fiber robotic Hectospec instrument \citep{fabricant05} on the 6.5m MMT
telescope. AGES obtained spectra for $\approx 26000$ objects in the
wavelength range from 3200 to 9200\AA. Spectroscopic redshifts have
been measured for about 18000 galaxies in the redshift range from 0 to
1 with a magnitude limit of $I=20$, and about 7000 likely AGNs in the
range from 0 to 6 up to $I=22.5$. Because of the different magnitude
limits for the samples, AGN candidates were pre-selected from the
photometric observations. A source was targeted as an AGN if it met
any of the following criteria:

\begin{itemize}
  \item {\it{X-Ray}}: The object was detected in X-rays by
  XBo\"otes. The contaminants in the X-ray sample are primarily active
  stars, along with a small number of very low redshift early and late
  type galaxies \citep[see e.g.][for a description of the origin of
  X-ray emission in galaxies]{stern02}.

  \item {\it{Radio}}: The object has an associated unresolved $5
  \sigma$ WSRT detection. Radio emission is generally associated with
  AGNs, but can also be due to star formation at low radio powers
  \citep[e.g.][]{hickox09,kauffmann08}. Strongly star forming galaxies
  at low redshift can contaminate this sample.

  \item {\it{MIPS}}: The object was detected in the 24$\mu$m MIPS
  images with a flux $F>0.3$~mJy, appears as a point source in the
  optical data and has an $I$-band magnitude fainter than $18 - 2.5
  \log(F/\rm mJy)$. This $I$-band magnitude limit eliminates normal
  stars. The MIPS sample is contaminated by star forming galaxies that
  are classified as point sources in the optical photometry.

  \item {\it{IRAC}}: The object has IRAC colors that suggest the
  presence of an AGN. The exact color-color boundaries and magnitude
  limits vary between optical point and extended sources that are
  discussed in detail in \citetalias{assef09}, but are based on
  modified versions of the \citet{stern05} criterion.

\end{itemize}

The final AGES sample includes spectra for 3047, 789, 2025 and 3844
objects targeted as AGN by the X-ray, radio, MIPS and IRAC criteria
respectively. 

After the observations, spectra were analyzed using a modified version
of the SDSS spectral pipeline. Based on the template fits to the
spectra, objects that showed clear signatures of nuclear activity
were classified as spectroscopic AGNs. This classification has limits,
as obscured AGN may lack optical signatures of activity, and the
pipeline classifications emphasize the presence of broad lines over
the more subtle classification problems for narrow lines. For narrow
line sources we also used the emission line classifications of
\citet{moustakas09}. We view these various AGN classifications as
complimentary rather than focusing only on one.

\begin{figure}
  \begin{center}
    \plotone{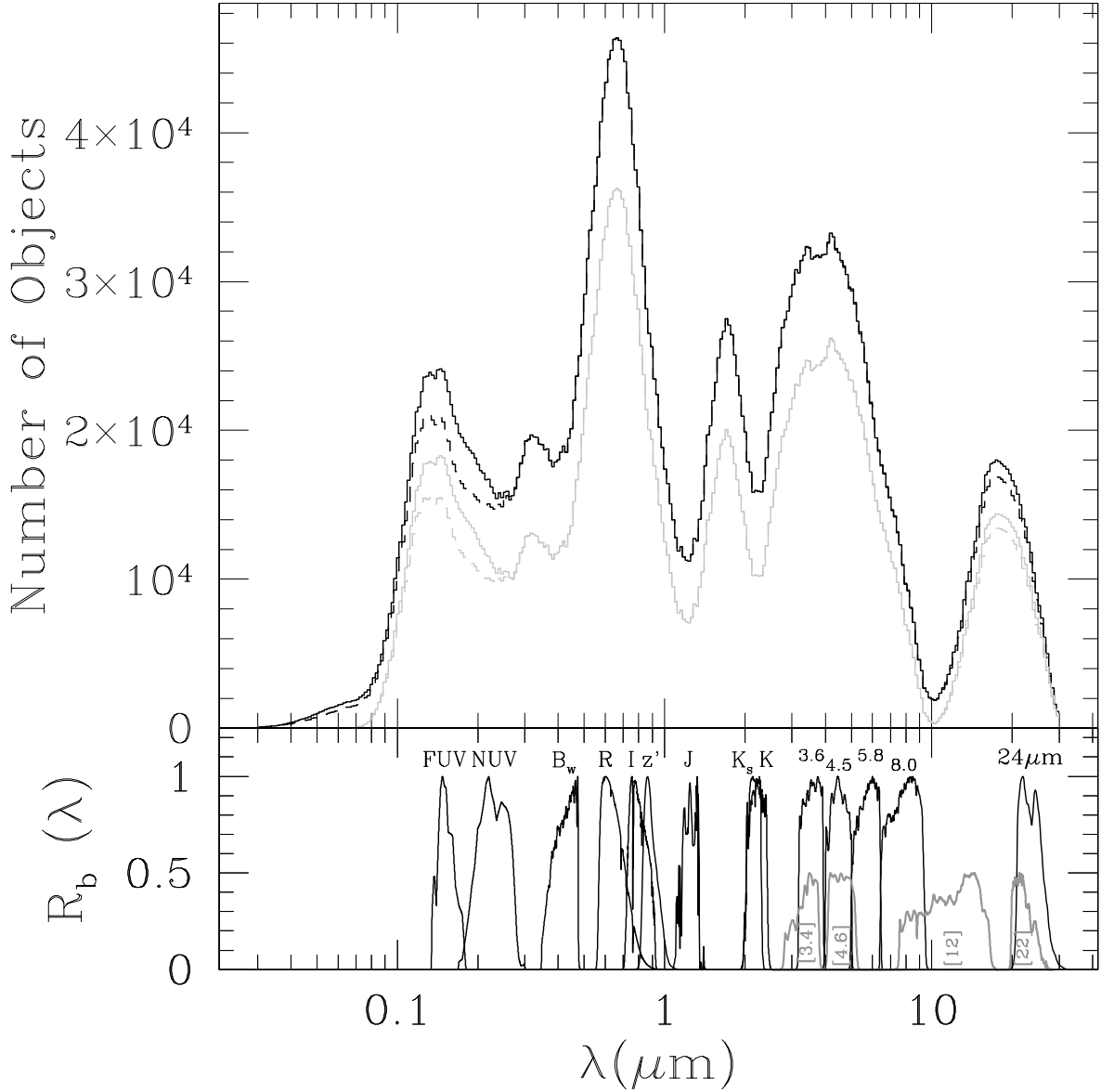}
    \caption{({\it{Upper panel}}) The wavelength coverage of our data
    set ({\it{black}}), as described in \S\ref{sec:data}. The solid
    line shows the coverage when photometry upper limits are included
    while the dashed line only considers detections. We also show both
    histograms for the sample used to construct the galaxy SED
    templates alone ({\it{gray}}) which excludes AGNs (see
    \S\ref{ssec:met_temps}). ({\it{Bottom panel}}) Filter sensitivity
    curves for all bands included in our data set are shown by the
    solid black lines. Filter sensitivity curves for the upcoming WISE
    mission are shown by the dark gray solid lines at half-height to
    distinguish them from the other bands.}
    \label{fg:wave_cont}
  \end{center}
\end{figure}

When possible, we have added upper limits to all objects observed, but
not detected, in the UV, optical, near-IR and 24$\mu$m. The catalog
from which we build the SED templates consists of all objects that
have been detected, or have upper limits, in at least 8 of the 14
bands available (FUV, NUV, $B_{W}$, $R$, $I$, $z'$, $J$, $Ks$, $K$,
[3.6], [4.5], [5.8], [8.0] and 24$\mu$m) and which have a reliable
redshift measured by AGES. We eliminate objects close to bright
stars. The minimum number of bands is required to guarantee that all
SED fits have at least 2 degrees of freedom (see
\S\ref{sec:methods}). Our final data set comprises 19795 objects, of
which 14448 are nominally ``pure'' galaxies with no signs of nuclear
activity and 5347 show some type of AGN signature. On average, each
object has been detected in 10 of the 14 available bands, and Figure
\ref{fg:wave_cont} shows the rest-frame wavelength coverage of our
final sample. 

\begin{figure}
  \begin{center}
    \plotone{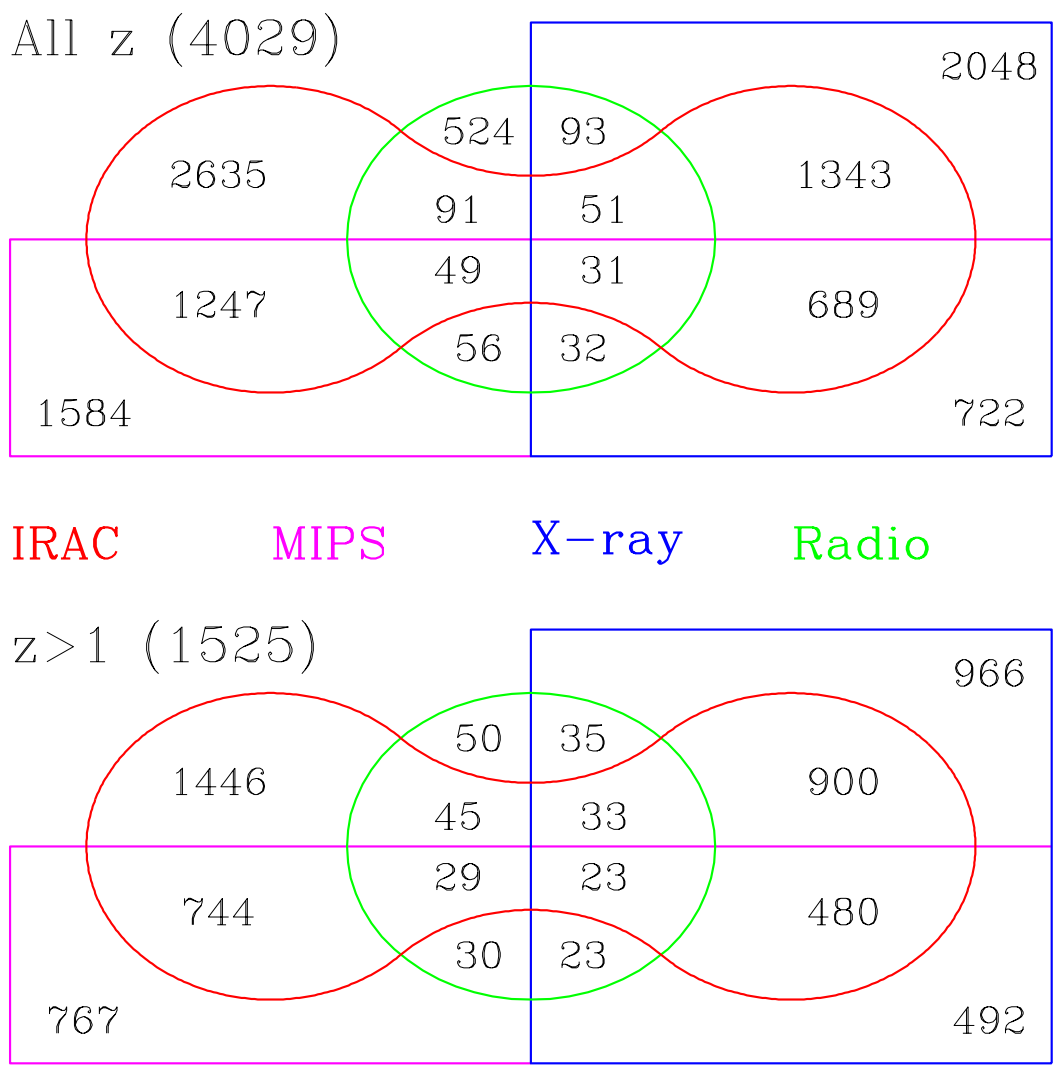}
    \caption{Edwards representation of a four set Venn diagram showing
    the number of AGNs selected by each photometric criteria and the
    overlap between them. The top panel shows this for our full sample
    while in the bottom panel we have restricted it to objects with
    $z>1$ in order to show how the photometric selection criteria
    overlap in a contamination free, but limited, sample. Note that in
    each panel there are four geometrical figures, each representing a
    photometric selection criteria: MIPS (magenta rectangle), X-ray
    (blue rectangle), radio (green circle) and IRAC (red ``peanut''
    shaped region). The intersection between the regions show the
    number of objects targeted simultaneously by each combination of
    criteria.}
    \label{fg:venn}
  \end{center}
\end{figure}

Figure \ref{fg:venn} shows a Venn diagram representation of the 4029
photometrically selected AGNs in our sample. Each geometrical shape
represents a different photometric selection criteria, and the
intersections between them show the number of objects targeted
simultaneously by each of those criteria. The rest of the AGNs in our
sample are spectroscopically confirmed by their line ratio
classification of \citet[1242 objects in total of which 1034 do not
meet any photometric targeting criteria]{moustakas09} or by the
analysis of the spectroscopic pipeline (2418 objects in total of which
268 were not targeted by their photometry and were not classified as
Type 2 AGNs by \citealt{moustakas09}). The remaining 16 active objects
were targeted by their optical colors using a selection technique of
minor importance in the AGES survey. Figure \ref{fg:venn} also shows
the Venn diagram representation of all $z>1$ objects, in order to show
how the photometric selection criteria overlap in the absence of
contaminants. Notice, however, that each selection criteria has
different redshift-dependent biases, so the differences between the
two Venn diagrams are not solely due to contaminating inactive
galaxies.

\section{Methods}\label{sec:methods}

The methods we use to derive the SED templates and estimate
photometric redshifts follow closely those described in
\citetalias{assef08}. In this section we give a brief summary of how
our methods work and we detail the changes we have made to the
original algorithms. We refer the reader to \citetalias{assef08} for a
detailed discussion of the latter.

\subsection{Templates}\label{ssec:met_temps}

We assume that the spectrum of any object in our sample can be modeled
as a linear combination of a small set of unknown spectral
templates. We construct these templates by using the data described in
\S\ref{sec:data}. For ``pure'' galaxies, we assume the majority of
them have SEDs that can be described as a linear combination of three
templates: one similar to an elliptical galaxy (an old stellar
population), one similar to a spiral galaxy (a continuously star
forming population), and a third similar to an irregular galaxy (a
starburst population). For objects with active nuclei, a population
that was not included in the analysis of \citetalias{assef08}, we
assume that every AGN SED can be described by the same spectral
template with varying amounts of reddening and absorption by the
intergalactic medium (IGM), combined with the galaxy templates to
describe the host. We assume the reddening follows an SMC-like
extinction curve for $\lambda < 3300$\AA\ and a Galactic extinction
curve at longer wavelengths, assuming $R_V = 3.1$ for both. This is
based on the observed absence of the Galactic 2175\AA\ feature of
Galactic extinction curves in QSO spectra \citep[e.g.][]{york06}. For
the SMC reddening law we assume the functional form derived by
\citet{gordon98} using the star AzV 18, while for the Galactic
reddening we assume the functional form of \citet{cardelli89}. The IGM
absorption is applied to all templates and not solely the AGN one, but
it is most important for the AGN due to their high redshifts.  We
model the IGM absorption following \citet{fan06} for Ly$\alpha$ and
Ly$\beta$ absorption, and \citet{stenglerlarrea95} for Lyman limit
systems. We leave the strength of absorption as a free parameter of
the model in order to allow for variations in absorption between
objects. We only include the IGM absorption component for redshifts
above 0.8, as below this redshift Ly$\alpha$ falls outside of our
wavelength coverage. To force every combination of these four
templates to be physically plausible, we require the SED of every
object to be a non-negative linear combination of these templates.

Our algorithm is iterative \citepalias[see discussion below and \S3.1
of][]{assef08} and hence we must start from an initial guess and
sequentially improve it to best fit the data at each iteration. For
the galaxy templates, we start from the \citet{cww} E, Sbc and Im
template for the ``elliptical'', ``spiral'' and ``irregular''
components, extended into the UV and IR by the \citet{bc03} stellar
synthesis models. Because the models of \citet{bc03} do not consider
dust or PAH emission, which are critical in the mid-infrared for the
star forming templates, we added ad-hoc linear combinations of the
mid-infrared part of the VCC1003 (NGC~4429) and M82 SEDs obtained by
\citet{devriendt99} to the Sbc and Im templates. For the AGN template
we consider two different starting points: an ad-hoc template
constructed from different observational and theoretical
considerations, and the average Type 1 quasar SED of
\citet{richards06a}. The ad-hoc template consists of a combination of
modified power-laws that resemble the large scale behavior of the
\citet{richards06a} Type 1 quasar mean energy distribution in the IR
and optical, modified in the UV to fall as predicted by a simple
accretion disk model, and with the addition of the broad-spectral
lines of an AGN. We choose to make our initial template as blue in the
optical as our bluest AGNs, which is significantly bluer than the
\citet{richards06a} SED. The shape of this template, along with the
initial guess for our galaxy templates, is shown in Figure
\ref{fg:temps}. Our results refer to this starting model except where
noted otherwise. While the two model templates produce very similar
results, we prefer the ad-hoc starting point because the
\citet{richards06a} SED has considerable amounts of host contamination
(see \S\ref{ssec:res_temps}).

Each template is divided into 300 uniformly logarithmically spaced
wavelength bins extending from 0.03 to 30$\mu$m, hence the ``low
resolution''. Each bin is then directly fit to the data using $\chi^2$
minimization over the entire calibration sample. Because this is still
a significantly higher resolution than the width of the broad-band
filters, we add a smoothing term ($H$) to the $\chi^2$ minimization
that penalizes departures from the smoothness of the initial
guesses. Note that this resolution is higher than that used in
\citetalias{assef08}. In summary, we optimize the function
\begin{equation}\label{eq:G}
G\ =\ \chi^2\ +\ \frac{1}{\eta^2} H.
\end{equation}
\noindent The $\chi^2$ term is given by
\begin{equation}
\chi^2\ = \sum_{i,b}\left(\frac{F_{i,b}-F^{\rm
mod}}{\sigma_{i,b}}\right)^2 ,
\end{equation}
\noindent where indices $i$ and $b$ enumerate objects and bands
respectively, $F$ is the observed flux, $F^{\rm mod}$ is the template
model flux and $\sigma_{i,b}$ is the photometric error. We defined the
smoothing term $H$ to be
\begin{equation}
H\ =\ \sum_{k,n} \left( \log{\frac{T_{k,\nu_n}}{Q_{k,\nu_n}}} -
\log{\frac{T_{k,\nu_{n+1}}}{Q_{k,\nu_{n+1}}}} \right)^2, 
\end{equation}
\noindent where $T_{k,\nu_n}$ is the flux per unit frequency of SED
template $k$ in wavelength bin $n$ (frequency $\nu_n$), and
$Q_{k,\nu_n}$ is the corresponding value in our initial template
guess. For a detailed discussion of these terms, we refer the reader
to \citetalias{assef08}. For galaxies, we use $\eta = 0.002$, which is
half of that used in \citetalias{assef08} due to the increased
resolution, but the results are not very sensitive to this choice. For
the AGN template we increased the value to $\eta = 0.005$ as the
amplitude of the changes is much smaller than for galaxies.

The equations for optimizing equation (\ref{eq:G}) are linear in all
parameters but the AGN reddening and the IGM absorption. To solve the
linear equations while imposing the condition that the contribution of
each template to each SED must be strictly non-negative and that at
every wavelength the templates must be positive, we use a simple
non-negative least square solver. 

To summarize, our algorithm proceeds as follows: (1) we estimate the
contribution of each template to the SED of each object; (2) we
estimate the zero point correction to each photometric band; (3) we
sequentially optimize the templates; (4) return to step (1). Note that
when optimizing a given template we only use objects with at least a
20\% contribution from the template to the integrated luminosity of
its best fit SED in the previous step. Since the 24$\mu$m data are
effectively decoupled from the rest of the data (see
Fig. \ref{fg:wave_cont}), we cannot fit a zero-point correction to
this channel. For this reason, unlike in \citetalias{assef08}, we do
not attempt to remove the large scale behavior of the zero point
corrections. Instead, we include a Gaussian prior to penalize
unnecessary modifications to the zero points, with a dispersion of 1\%
for all bands except the MIPS 24$\mu$m channel.

We first derive the best fit galaxy SED templates without the AGN
template by using only objects that show no signatures of nuclear
activity. To be conservative, we eliminate all objects photometrically
targeted as AGNs by AGES, all objects for which the SDSS pipeline
finds an AGN component in the spectrum, and all objects classified as
narrow line AGNs by \citet{moustakas09} (see \S\ref{sec:data}). We
then fit the AGN template by using our full sample while keeping the
galaxy templates fixed. When determining the galaxy templates, we
eliminate the 3\% of the objects with the poorest fits, while for
determining the AGN template we eliminate the worst 5\%. These are
mostly objects with either bad data points or galaxies with unflagged
nuclear activity.

\subsection{Photometric Redshifts}\label{ssec:met_zphot}

For photometric redshifts we follow the same basic steps as those in
\citetalias{assef08}. We perform a $\chi^2$ minimization similar to
that outlined in \S\ref{ssec:met_temps}, solving all linear equations
with the NNLS algorithm combined with a grid of reddening values for
the AGN template as a function of redshift. The main difference is
that we fix the IGM absorption to a typical value rather than letting
it vary (see \S\ref{ssec:res_zphot} for a discussion of this
point). For optimizing the photometric redshifts, we also include a
luminosity prior based on the Las Campanas Redshift Survey
\citep{lcrslf} luminosity function that only affects the galaxy
templates (with and without and AGN component on the SED). The shape
and specific parameters of the prior probability function are discussed
in detail in \citetalias{assef08}. Our publicly available
algorithms\footnote{Codes and templates are available at
\url{www.astronomy.ohio-state.edu/~rjassef/lrt}} for estimating
photometric redshifts have been updated to reflect these changes and
include the new templates.

\section{Results}\label{sec:results}

In this section we describe the SED templates resulting from applying
the algorithm discussed in \S\ref{ssec:met_temps} to the data set
described in \S\ref{sec:data} and we also discuss their application to
determining photometric redshifts for AGNs and galaxies, and to study
the mid-IR selection of QSOs in the IRAC bands and in the upcoming
WISE satellite mission.

\subsection{Templates}\label{ssec:res_temps}

\begin{figure}
  \begin{center}
    \plotone{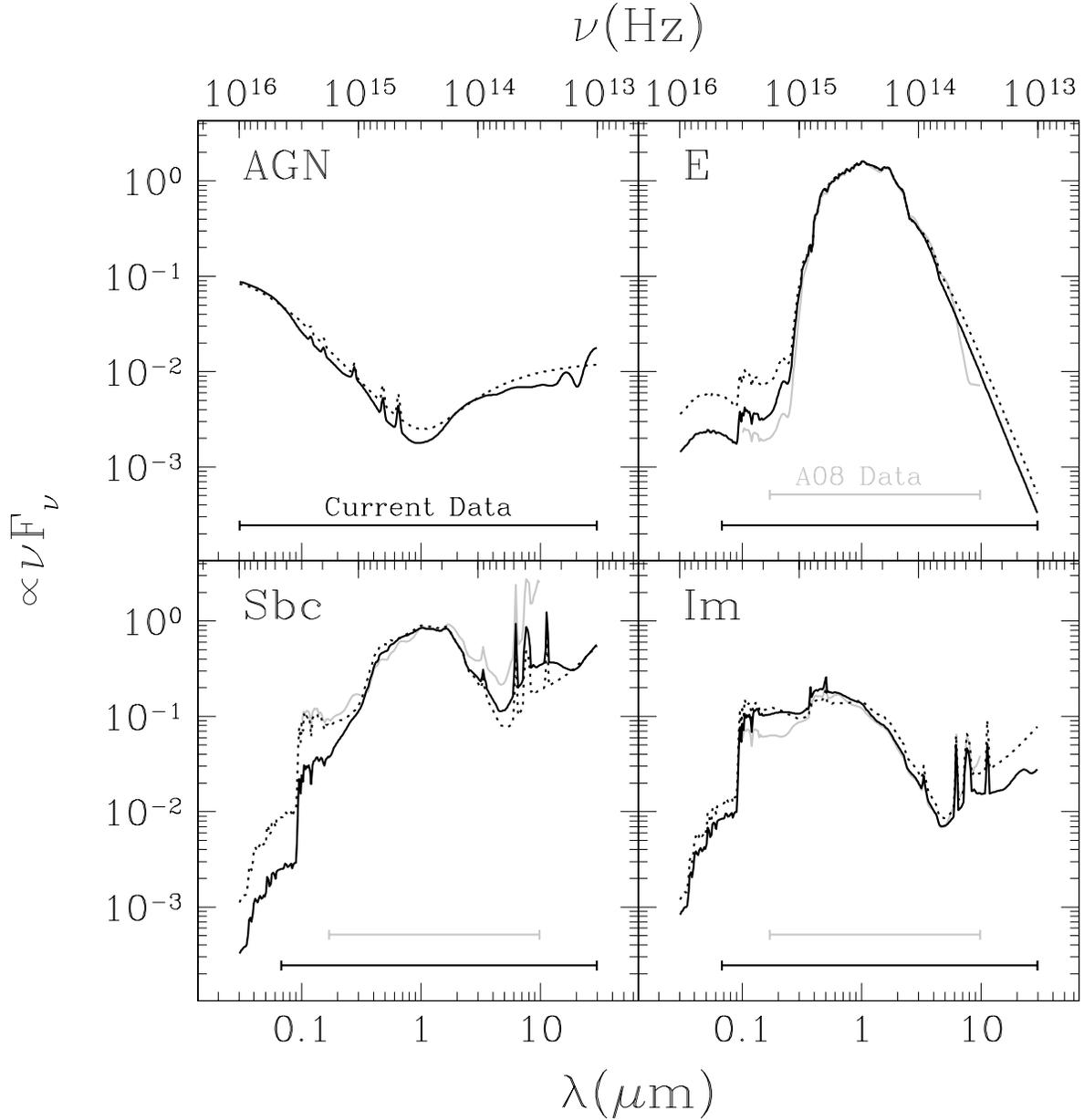}
    \caption{Resulting best fit templates from applying the algorithm
    of \S\ref{ssec:met_temps} to the data set described in
    \S\ref{sec:data} ({\it{black solid}}) compared to their initial
    guesses ({\it{black dotted}}) and to the templates derived in
    \citetalias{assef08} ({\it{gray solid)}}. Note that the bottom
    axis of each panel shows wavelength while the top shows
    frequency. The bars at the bottom of each panel show the
    rest-frame wavelength coverage of our data set ({\it{black}}, see
    Fig. \ref{fg:wave_cont}) and of that used in \citetalias{assef08}
    ({\it{gray}}).}
    \label{fg:temps}
  \end{center}
\end{figure}

The final templates are shown in Table \ref{tab:spectab} and in Figure
\ref{fg:temps}. Table \ref{tab:kcorr} shows the absolute magnitudes of
the templates as a function of redshift in various photometric bands
including those of our sample. Figure \ref{fg:temps} also shows the
initial guesses from which the process started. The galaxy templates
show significant departures from the initial ones, just as in
\citetalias{assef08}. In particular, the elliptical template is
significantly redder in the UV and somewhat bluer in the mid-IR. The
spiral-like template has decreased UV and a increased dust/PAH
emission relative to the 1.6$\mu$m stellar emission peak. The Im
template shows a somewhat decreased UV and mid-IR continuum relative
to the near-IR peak. The departures from the initial templates are
caused by a combination of changes in the balance of the underlying
stellar populations and by their inability to reproduce the SEDs of
galaxies, either because of missing physics or, in the mid-IR, because
of the ad hoc addition of dust emission used to create them (see
\S\ref{ssec:met_temps} for details).

The AGN template based on our ad-hoc model also shows significant
differences from the initial guess. In particular, the optical is
bluer and the mid-IR is redder. At longer wavelengths ($\gtrsim
15\mu\rm m$), the template shows a steep rise, but this is likely to
be an artifact caused by the lack of $z=0$ quasars to constrain this
wavelength region. At wavelengths shortward of Ly$\alpha$, the
templates cannot be constrained, as the observed SEDs are completely
dominated by IGM absorption. Since the flux in X-rays is generally
significantly lower than the UV emission in AGNs, as determined by the
values of $\alpha_{OX}$ \citep[e.g.][]{tananbaum79,strateva05}, we
know the spectrum must turn down at some wavelength shortward of
Ly$\alpha$. For this reason, we choose in the rest of the paper to
estimate ``bolometric'' luminosities by integrating our AGN template
only at wavelengths longward of Ly$\alpha$.

\begin{figure}
  \begin{center}
    \plotone{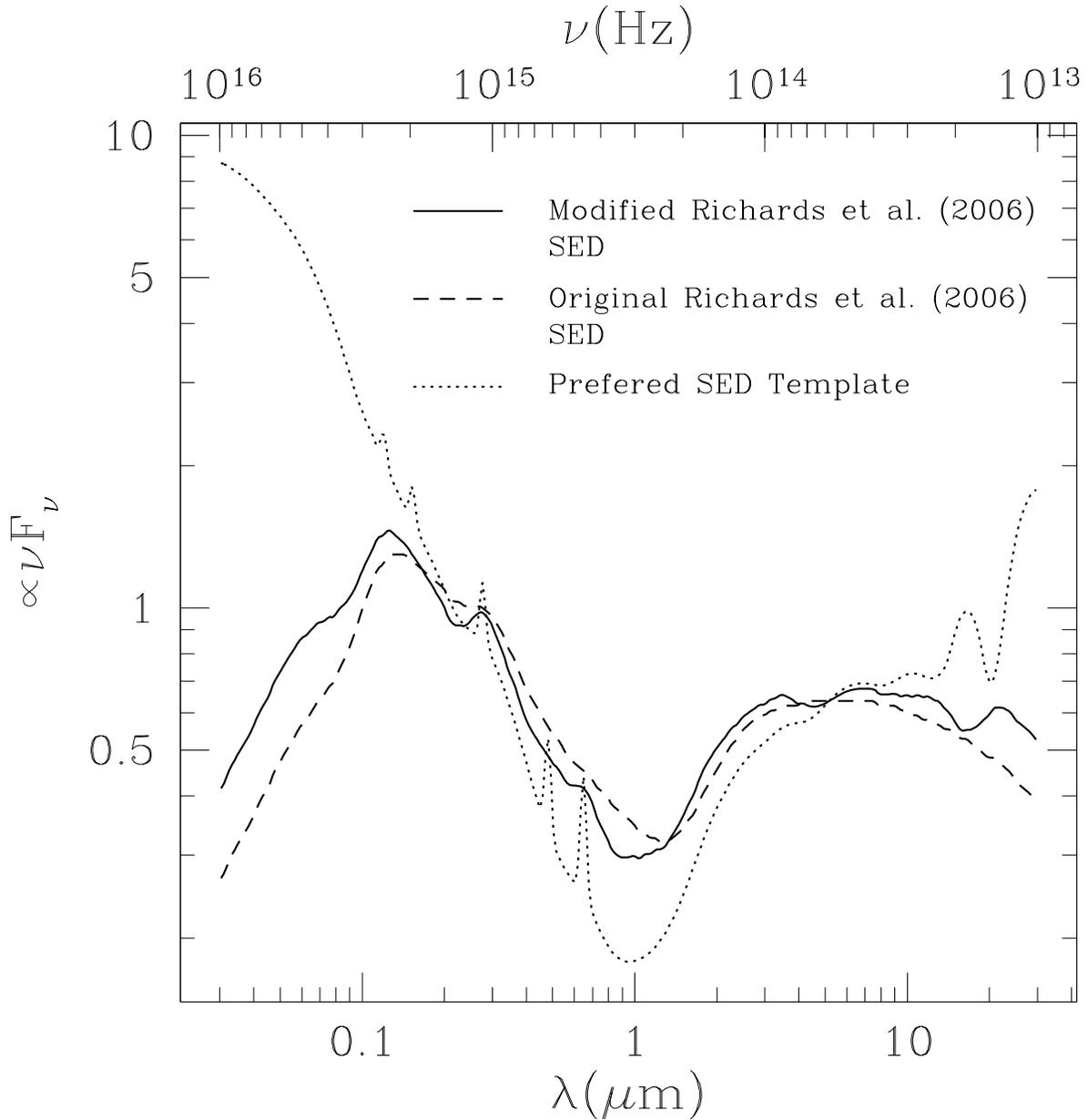}
    \caption{Converged AGN SED template using the \citet{richards06a}
    template as a starting point ({\it{solid line}}) and our standard
    model ({\it{dotted line}}). The Figure also shows the original SED
    template of \citet{richards06a} ({\it{dashed line}}). The
    templates are poorly constrained near $\gtrsim 20\mu\rm m$ due to
    the lack of powerful low redshift quasars.Note that the bottom
    axis shows wavelength while the top shows frequency.}
    \label{fg:richards_med}
  \end{center}
\end{figure}

\begin{figure}
  \begin{center}
    \plotone{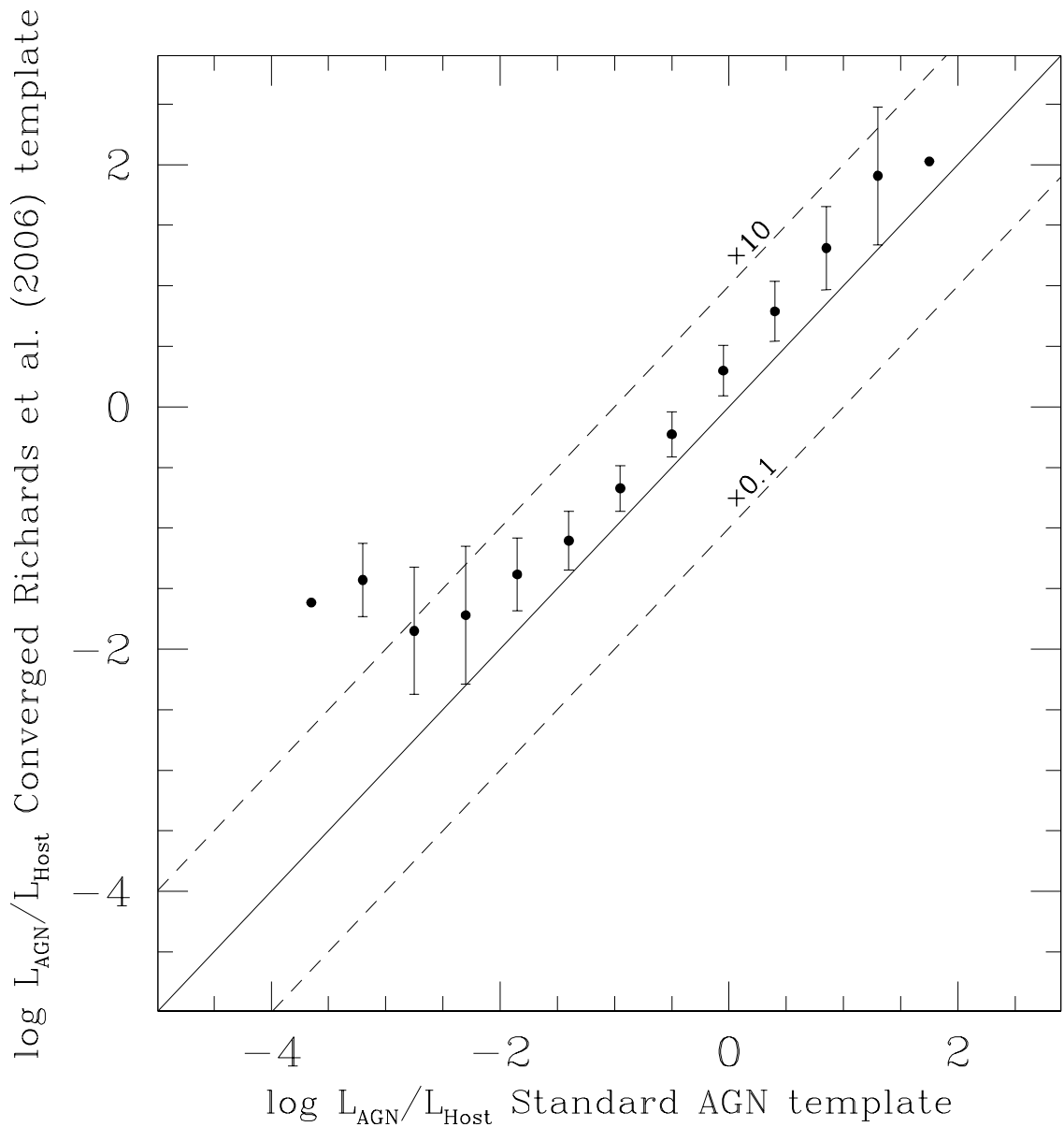}
    \caption{Average $L_{\rm AGN}/L_{\rm Host}$ obtained using the
    converged \citet{richards06a} AGN SED template as a function of
    the ratio found using our standard AGN template. On the solid line
    the ratios are equal, while on the dashed lines they differ by a
    factor of 10. Error bars show the dispersion in each bin of
    $L_{\rm AGN}/L_{\rm Host}$ determined with the standard AGN
    template.}
    \label{fg:comp_ratios_seds}
  \end{center}
\end{figure}

Using the average SED of Type 1 quasars derived by \citet{richards06a}
as a starting point for our AGN template yields similar results,
albeit with some interesting differences. The AGN template resulting
from this starting point is shown in Figure \ref{fg:richards_med}, as
compared to the original \citet{richards06a} template and our
``standard'' template. Relative to the initial template, the converged
template is somewhat brighter in the UV relative to the optical and
near-IR wavelengths. The mean color is a combination of extinction and
any dependence of the SED on black hole mass and accretion rate
\citep[e.g.][]{yip04}. While our algorithm corrects for the former, it
can mimic the latter using small amounts of reddening. In the optical
wavelengths, the contribution from broad emission lines to the
\citet{richards06a} template is heavily smoothed by the averaging of
the broad band photometry used to build it, however the output
template to our algorithm has developed much sharper emission line
features. The minimum in the near-IR of the converged template has
shifted to a shorter wavelength, as also seen in our ``standard''
template. This is likely due to our self-consistent treatment of the
host. At $\sim4.5\mu$m there is a depression in the SED. It is unclear
if this is a real feature or not, as it does not appear in our
standard AGN template. The differences at the longest wavelengths and
wavelengths shorter than Ly$\alpha$ are not significant due to the
lack of data needed to constrain these regions (see
Fig. \ref{fg:wave_cont} and the discussion in the previous
paragraph). When combined with the host galaxy templates, both AGN
templates work well, suggesting that the main difference is that the
\citet{richards06a} template intrinsically has more host contamination
than our standard template, which can be compensated for by changing
the amount of host galaxy template added to fit an object. This can be
seen in Figure \ref{fg:comp_ratios_seds}, which shows that the values
of $L_{\rm AGN}/L_{\rm Host}$ are, in average, $\approx 35\%$ larger
for the converged \citet{richards06a} AGN template than for the
standard template.

\begin{figure}
  \begin{center}
    \plotone{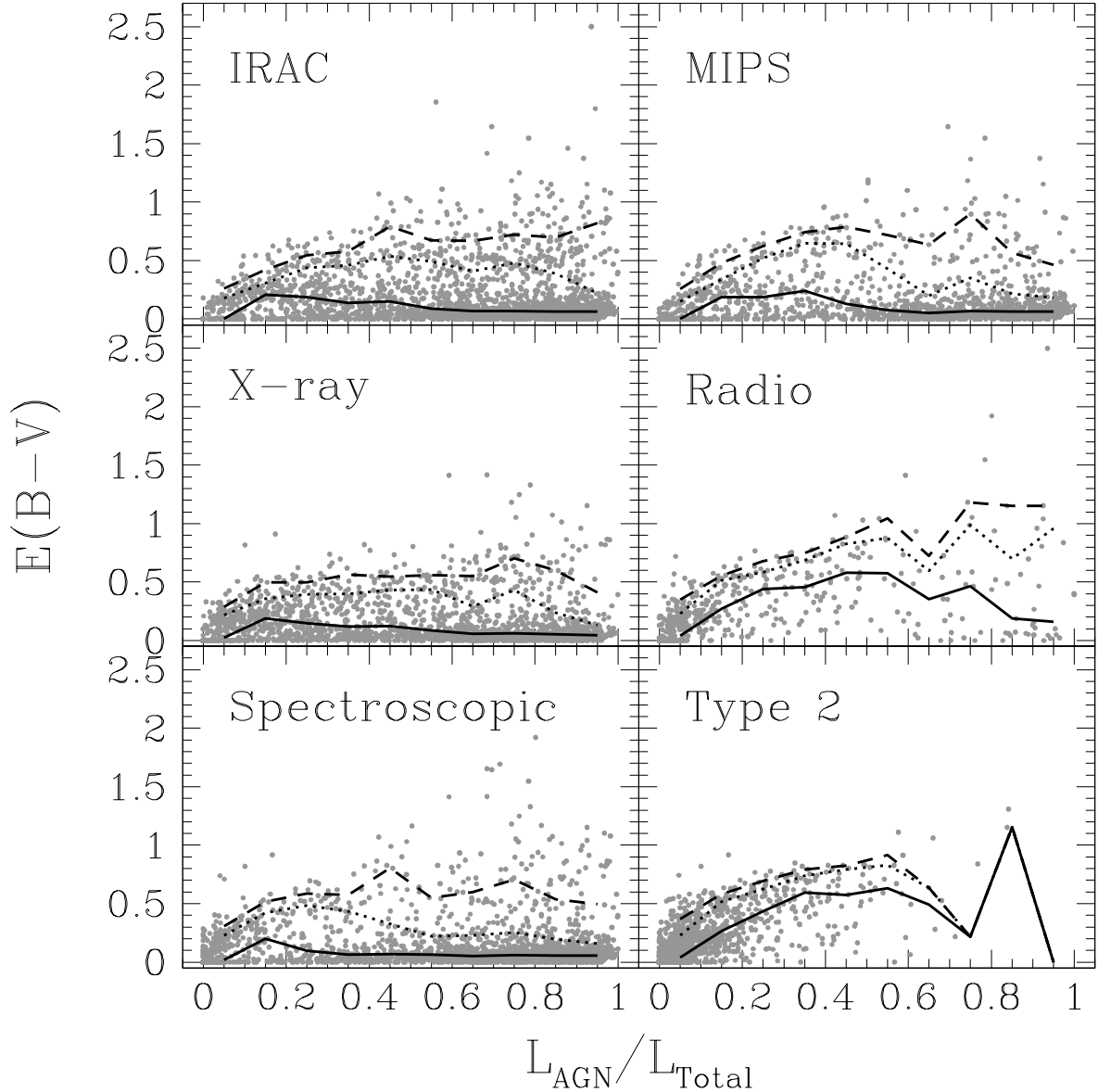}
    \caption{Best fit reddening to the AGNs in our sample as a
    function of the contribution of the AGN component to the total
    bolometric luminosity. The AGN luminosity has been corrected for
    the estimated extinction. The top and middle panels show the
    objects targeted as AGN candidates by AGES according to their
    photometric properties (see \S\ref{sec:data}). The bottom left
    panel shows the reddening distribution for objects classified as
    AGNs by the spectroscopic pipeline while the bottom right panel
    shows it for the objects classified as Type 2 AGNs by
    \citet{moustakas09}. In each panel, the solid black line shows the
    median E(B--V) as a function of $L_{\rm AGN}/L_{\rm Host}$ while
    the dotted and dashed lines show the 85\% and 95\% contours
    respectively.}
    \label{fg:ebv}
  \end{center}
\end{figure}

In Figure \ref{fg:ebv} we show the best fit extinctions to the AGNs in
our sample for our standard AGN template. Objects have been divided
according to the criteria by which they were selected and objects
targeted by multiple criteria are shown in all of the relevant
panels. In all cases, there seems to be a maximum observed reddening
that increases with increasing AGN luminosity relative to their
hosts. This can be understood as a combination of three effects. One
is that a more heavily reddened AGN needs to be more luminous relative
to the host in order to meet many of the AGES AGN targeting
criteria. The second effect is the inability of our templates to
detect highly reddened AGNs with bright hosts and, if detected, to
determine the reddening parameter with precision. If the host
dominates the UV/optical/NIR wavelengths, reddening values above a
certain threshold will not significantly affect the $\chi^2$ until
they start to modify the mid-IR colors. Otherwise, our algorithm
defaults to using the smallest possible reddening value. Finally, our
sample is magnitude limited, and hence the more reddened, optically
fainter objects can be found in a smaller physical volume than the
less reddened, optically brighter ones. The first problem primarily
affects the IRAC and MIPS targeting criteria, while the second and
third ones affect all cases. Notice, however, that the X-ray selected
objects are relatively unaffected by these biases. In Figure
\ref{fg:ebv} we also see that the AGNs which are brightest in
comparison to their hosts show a range of small, but positive,
extinctions. This may be due to our algorithm using small amounts of
reddening to compensate for changes in the SED with BH accretion rates
and masses, as discussed above.

Figure \ref{fg:temps} also shows the galaxy templates derived in
\citetalias{assef08}. The elliptical and irregular templates show
significant changes only in the UV and mid-IR, as would be expected
from the addition of the GALEX and $24 \mu$m MIPS data. The spiral
template also shows departures at those wavelengths, but differences
are also seen in the near- to mid-IR region. In our model the
templates form a non-orthogonal basis that, due to the restriction to
positive combinations, determine the boundaries of the color space
occupied by galaxies. Changes in the templates are really changes in
the boundaries of the permitted color space. We expect that the
differences between the new and old templates over the wavelength
range covered by the old templates can largely be explained by making
the new/old templates linear combinations of the old/new
templates. Note that the coefficients of the ``rotation'' can be
negative. The best fit combination of our new templates to the E
template of \citetalias{assef08} is given by the coefficients (1.02,
--0.01, 0.00), to the Sbc one by (--0.51, 1.51, 0.00) and to the Im
one by (--0.04, 0.34 and 0.70), where each coefficient represents the
contribution of our new E, Sbc and Im templates respectively. These
combinations reproduce very well the full wavelength range of the
\citetalias{assef08} templates, showing that the differences with our
new templates correspond to a simple rotation of the basis vectors
rather than a fundamental modification.

\begin{figure}
  \begin{center}
    \plotone{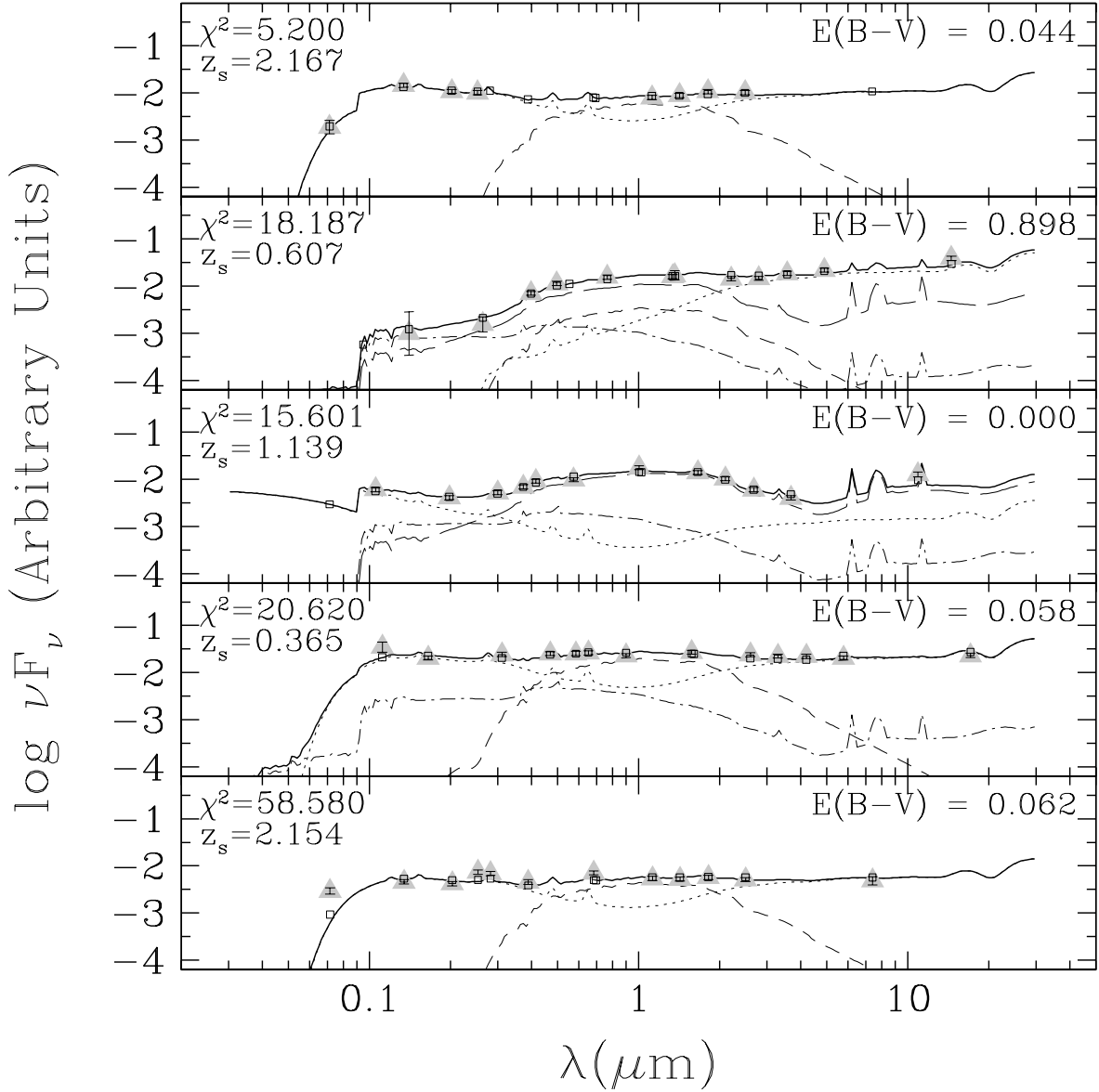}
    \caption{Best fit SEDs for four typical AGNs ({\it{top four
    panels}}) and for the AGN with the 90$^{\rm th}$ percentile worst
    $\chi^2_{\nu}$ in the sample ({\it{bottom panel}}). The solid line
    shows the best fit SED, which is composed of a non-negative
    combination of our optimized AGN ({\it{dotted}}), E
    ({\it{dashed}}), Sbc ({\it{long-dashed}}) and Im
    ({\it{dot-dashed}}) templates. In each panel we indicate the
    reddening applied to the AGN template. The flux of each filter,
    when detected, is shown by the solid triangles, while the open
    squares show the expectation from the best-fit. Arrows mark upper
    limits when they are available. The AGNs shown in the top four
    panels have SEDs typical of, from top to bottom, a Type 1, a Type
    2, a very low luminosity AGN and a Type 1 with a very strong host
    component. For each object, the $\chi^2$ of the fit and the
    spectroscopic redshift measured by AGES is shown on the upper
    right corner of the corresponding panel.}
    \label{fg:agn_fits}
  \end{center}
\end{figure}

\begin{figure}
  \begin{center}
    \plotone{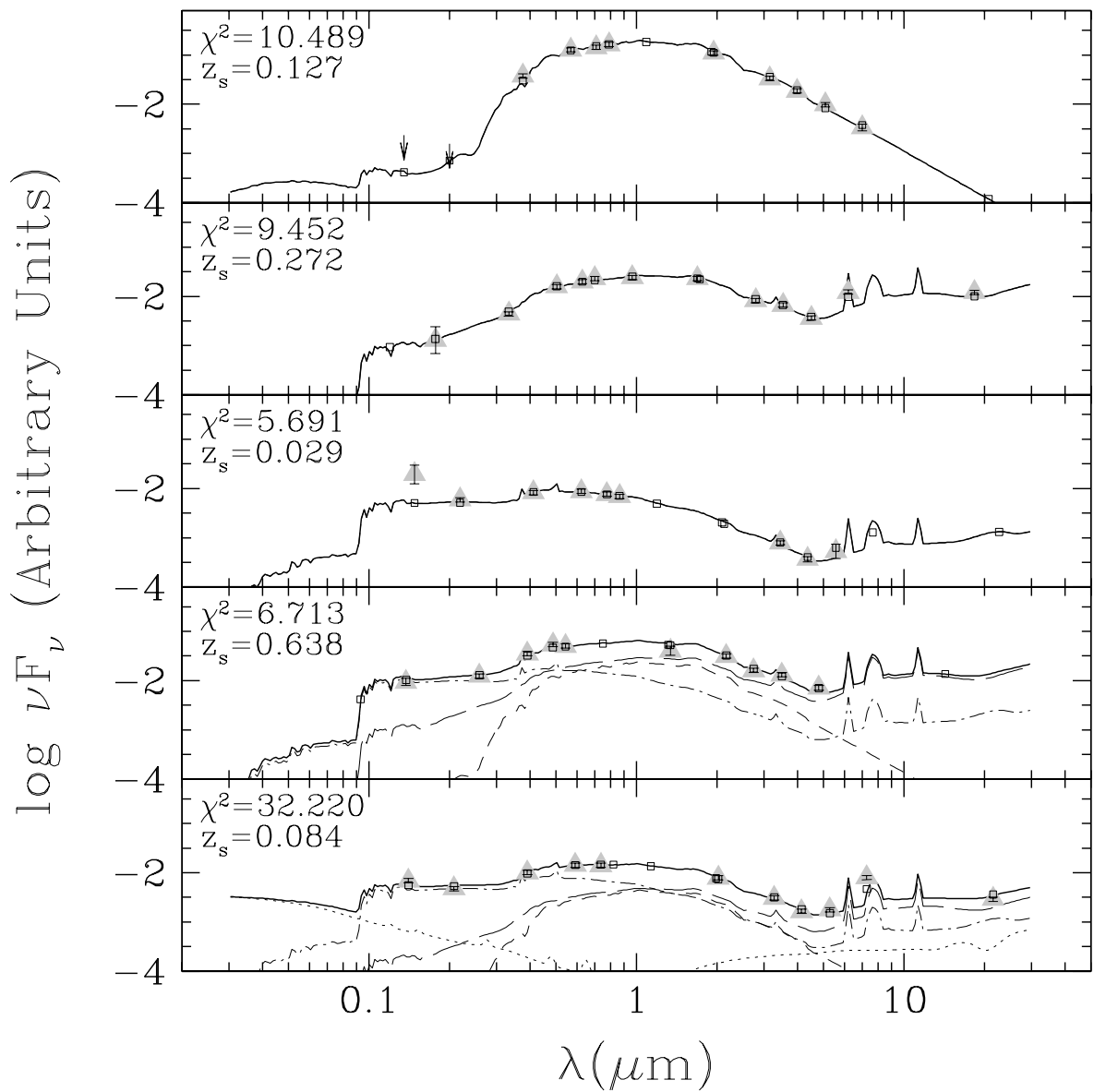}
    \caption{Best fit SEDs for four typical galaxies ({\it{top four
    panels}}) and for the galaxy with the 90$^{\rm th}$ worse
    $\chi^2_{\nu}$ in the sample ({\it{bottom panel}}). The different
    line styles and point types follow the same convention as in
    Figure \ref{fg:agn_fits}. In the top four panels we show galaxies
    dominated by, from top to bottom, the elliptical, Sbc and Im
    templates, and a combination of all three.}
    \label{fg:gal_fits}
  \end{center}
\end{figure}

Figures \ref{fg:agn_fits} and \ref{fg:gal_fits} show typical fits to
AGNs and galaxies using the ``standard'' AGN template. These objects
have $\chi^2$ values close to the median of their respective samples
(median $\chi^2_{\nu} = 1.35$ for galaxies and 2.60 for AGNs) and have
been selected to show a range of different scenarios. The bottom panel
of each Figure also shows the object with the 90$^{\rm th}$ percentile
worst $\chi^2_{\nu}$ of each sample. Although small discrepancies are
present at some wavelengths, the agreement is still generally very
good. The vast majority of the objects in our sample are well
described by our four templates. The converged \citet{richards06a}
template gives similar values of median $\chi^2_{\nu}$ (1.30 and 2.61
for galaxies and AGNs respectively) to our standard template as we
would hope, given that our results should depend little on the
priors.

In the following sections, we will explore different applications of
our templates. We will study the accuracy of photometric redshifts
from these templates for galaxies and AGNs, and the colors of these
objects in the mid-IR bands of both IRAC and the upcoming WISE
mission. In a subsequent paper \citepalias{assef09} we examine the AGN
luminosity function and its evolution with redshift. Our templates
also allow for the study of host properties in AGNs but this will be
explored in an upcoming paper.

\subsection{Photometric Redshifts}\label{ssec:res_zphot}

In this section we study the application of these templates to the
estimation of photometric redshifts for galaxies and AGNs. To quantify
their accuracy, we derive photometric redshifts for all the sources in
the catalog used to build the templates, each of which has a measured
spectroscopic redshift. As mentioned previously, for photo-zs we do
not allow the IGM absorption strength parameter to vary, as allowing
variations worsens the redshift accuracy, and we include a
luminosity prior based on the Las Campanas Redshift Survey $r$-band
luminosity function \citep{lcrslf}.

We divide our catalog into three distinct groups: ``pure'' galaxies,
extended AGNs (optically resolved galaxies with AGN signatures), and
point source AGNs (objects where the host galaxy was not optically
resolved), based on the AGES targeting criteria (see
\S\ref{sec:data}). The difference between the two AGN groups is a
combination of redshift and the relative brightness of the host and
the AGN. Photometric redshifts for all groups are estimated in the
same way, using all four templates and using a photometric redshift
range of $0\leq z \leq 6$. To study the accuracy of the
photometric redshifts, we calculate (i) the standard dispersion
$\sigma_z/(1+z)$ defined as
\begin{equation}\label{eq:sigz}
\left[\frac{\sigma_z}{(1+z)}\right]^2\ =\ \frac{1}{N} \sum_{i=1}^N
\left(\frac{z_p^i - z_s^i}{1+z_s^i}\right)^2 ,
\end{equation}
\noindent where $z_p$ and $z_s$ are the photometric and spectroscopic
redshifts respectively, (ii) the median offset of $z_p-z_s$, (iii) the
ranges of $|z_p-z_s|/(1+z_s)$ encompassing 68.3\%, 95.5\% and 99.7\%
of the distribution, and (iv) the dispersion $\Delta z$ which is
equivalent to $\sigma_z/(1+z)$ but uses only the 95\% objects with the
most accurate photometric redshift estimates in order to eliminate
outliers.

We note that including the MIPS 24$\mu$m data slightly worsens the
photometric redshift accuracies for all groups. This is in contrast to
including the mid-IR data, which improves the accuracy. Depending on
the object type, including the MIPS data increases the dispersion (as
measured by Eqn. [\ref{eq:sigz}]) by factors between 1.07 (galaxies)
and 1.21 (point source AGNs), and the 95\% clipped accuracy by between
1.05 (extended AGNs) and 1.09 (point source AGNs). This is probably
symptomatic of a scatter in the observed SEDs larger than permitted by
our templates, even when they fit the mean sample characteristics
well. This may be caused by a small number of LIRGs and ULIRGs that
should be modeled as a different population at these wavelengths. For
this reason we exclude the MIPS photometry when estimating photometric
redshifts for the rest of the paper.

\begin{figure}
  \begin{center}
    \plotone{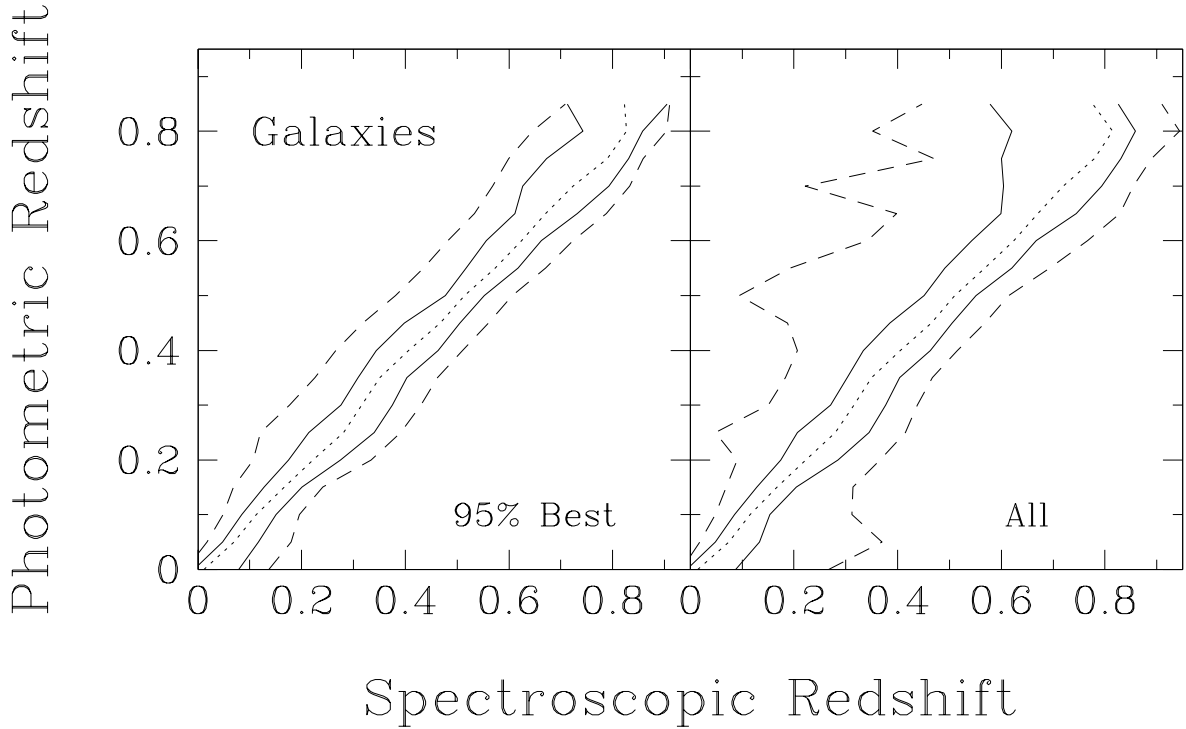}
    \caption{Comparison of photometric and spectroscopic redshifts for
    all the galaxies in our sample ({\it{right}}) and for the 95\%
    with the most accurate redshift estimates ({\it{left}}). For a
    given value of the photometric redshift, we show the median
    spectroscopic redshift of a galaxy with that $z_p$ estimate
    ({\it{dotted line}}), and the 68.3\% ({\it{solid line}}) and
    95.4\% ({\it{dashed line}}) range of the spectroscopic redshift
    distribution. We only show $z_p$ values for which there were
    enough objects to determine the 68.3\% and 95.4\% limits on both
    sides of the median.}
    \label{fg:zphot_gals}
  \end{center}
\end{figure}

\begin{figure}
  \begin{center}
    \plotone{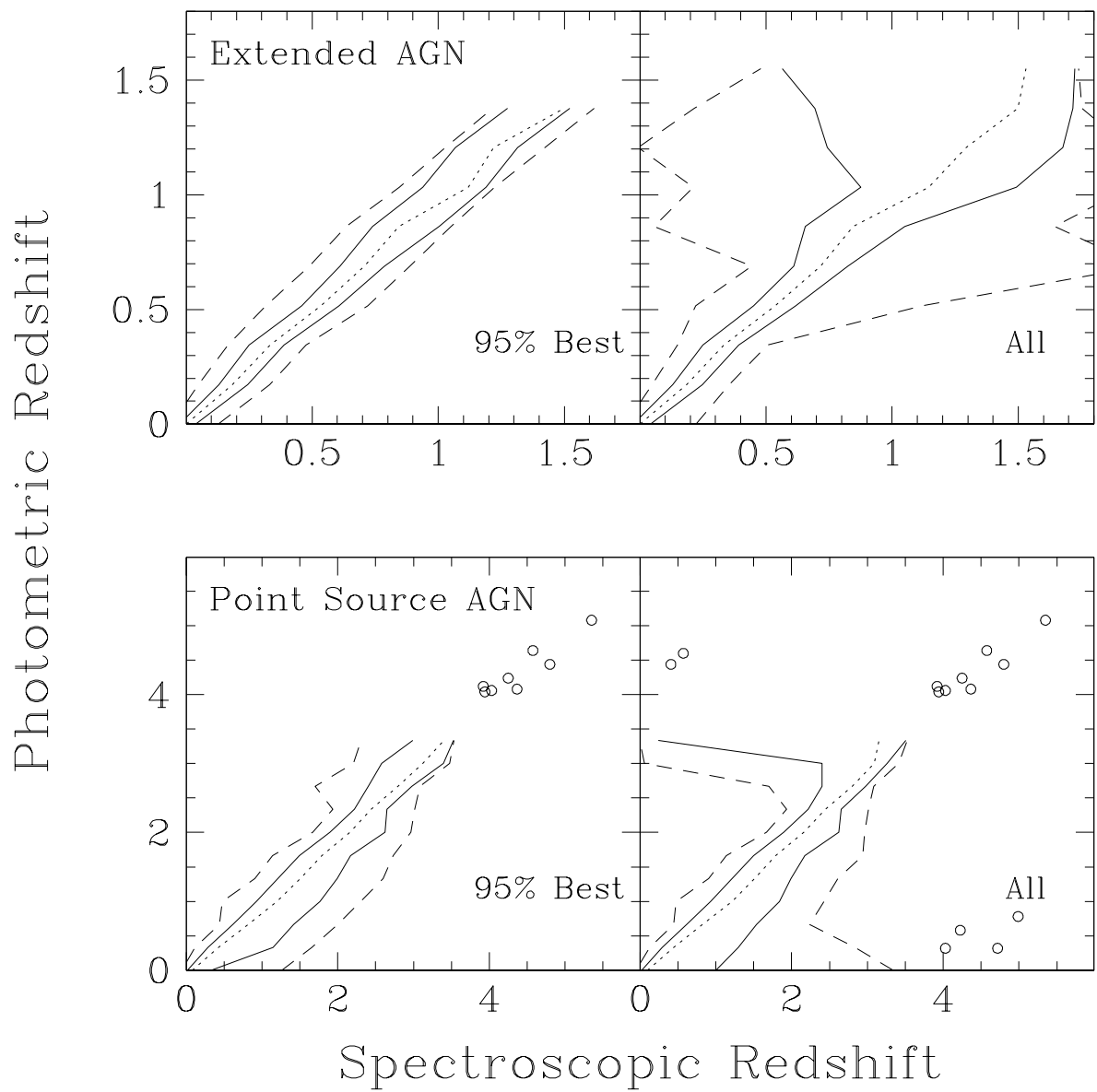}
    \caption{Comparison of photometric and spectroscopic redshifts for
    the extended ({\it{top}}) and point source AGNs ({\it{bottom}}) in
    our sample considering all ({\it{right}}) and the 95\% with the
    most accurate redshift estimates ({\it{left}}). Different line
    styles have the same definition as in Figure
    \ref{fg:zphot_gals}. The open circles show individual objects with
    either $z_s>4$ or $z_p>4$.}
    \label{fg:zphot_agns}
  \end{center}
\end{figure}

The results of our photometric redshift estimates are summarized in
Table \ref{tab:photoz}. Figure \ref{fg:zphot_gals} shows a comparison
of $z_p$ against $z_s$ for the galaxy sample, while Figure
\ref{fg:zphot_agns} shows it for the extended and point source
AGNs. For galaxies, the contours closely follow the $z_p = z_s$ line,
showing that our photometric redshifts estimates are accurate and that
most galaxies can indeed be described by our 4 template model. The
95\% clipped redshift accuracy is $\Delta z = 0.041$. These results
are compatible with those presented in \citetalias{assef08}, although
the accuracy is slightly worse. This is largely due to expanding the
redshift search range from 0--1 to 0--6 in order to encompass the AGN
population, even when it is not physical to do so. It is worth noting
that this accuracy is comparable to that found by \citet{brodwin06}
for galaxies in the NDWFS Bo\"otes field using a hybrid template
fitting and neural network approach and an earlier version of the AGES
data. \citet{brodwin06} found $\sigma_z/(1+z) = 0.105$ and $\Delta z =
0.047$, about 20\% better and 15\% worse, respectively, than the
accuracy we find.

\begin{figure}
  \begin{center}
    \plotone{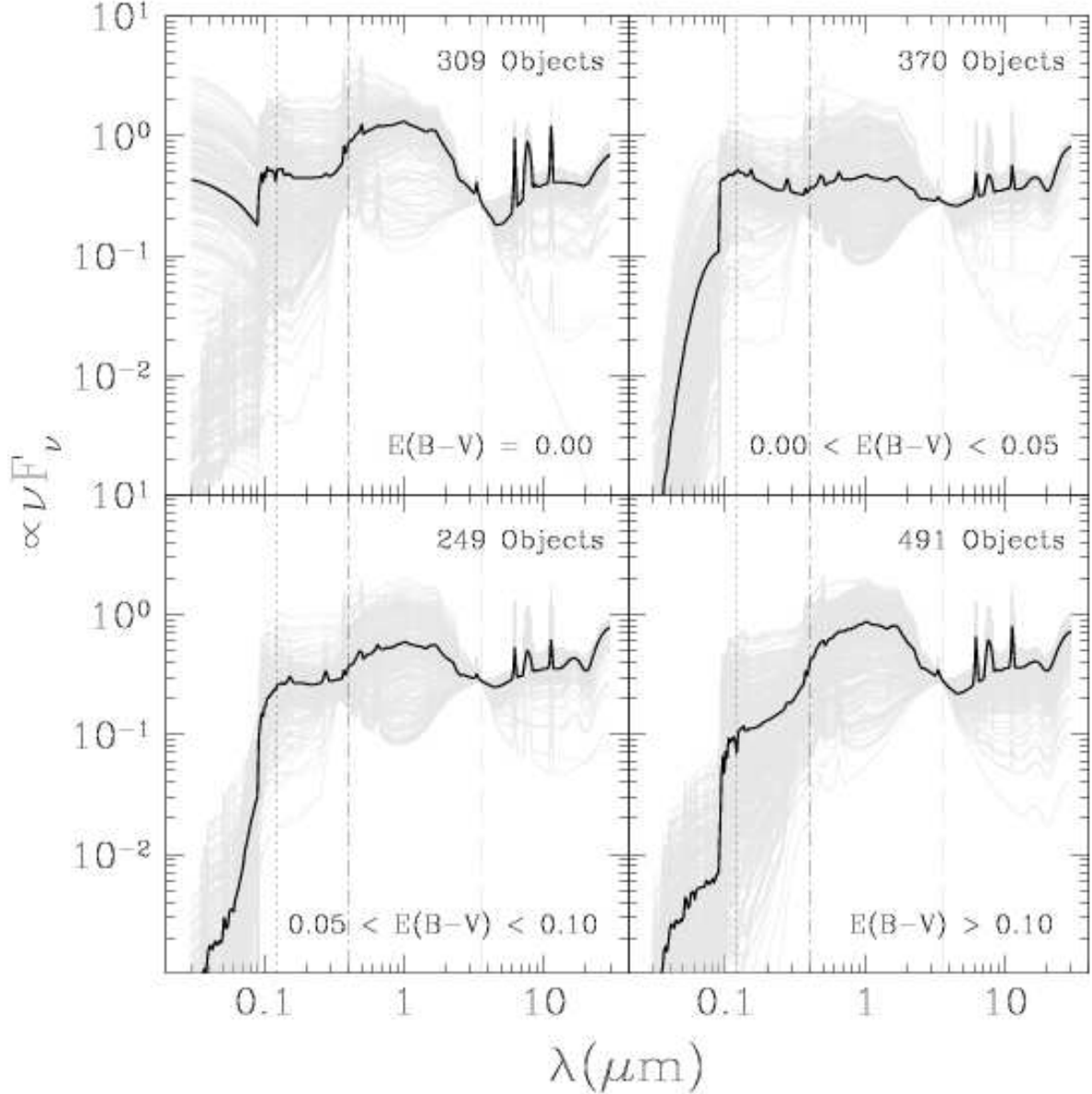}
    \caption{All best fit SEDs ({\it{gray}}) and their average
    ({\it{black}}) for point source AGNs with good photometric
    redshifts ($|z_p-z_s|<0.2$) in four ranges of the AGN
    reddening. The number of objects used in each case is shown in the
    top-right corner of each panel. All SEDs are normalized to match
    at 3.6$\mu$m. The vertical lines show the wavelength of
    Lyman-$\alpha$ ({\it{dotted}}), 4000\AA\ ({\it{dot-dashed}}) and
    3.6$\mu$m ({\it{dashed}}). The extreme scatter at wavelengths
    shorter than Ly$\alpha$ is due to large variations in IGM
    absorption with redshift.}
    \label{fg:SEDs_photo_z_good}    
  \end{center}
\end{figure}

\begin{figure}
  \begin{center}
    \plotone{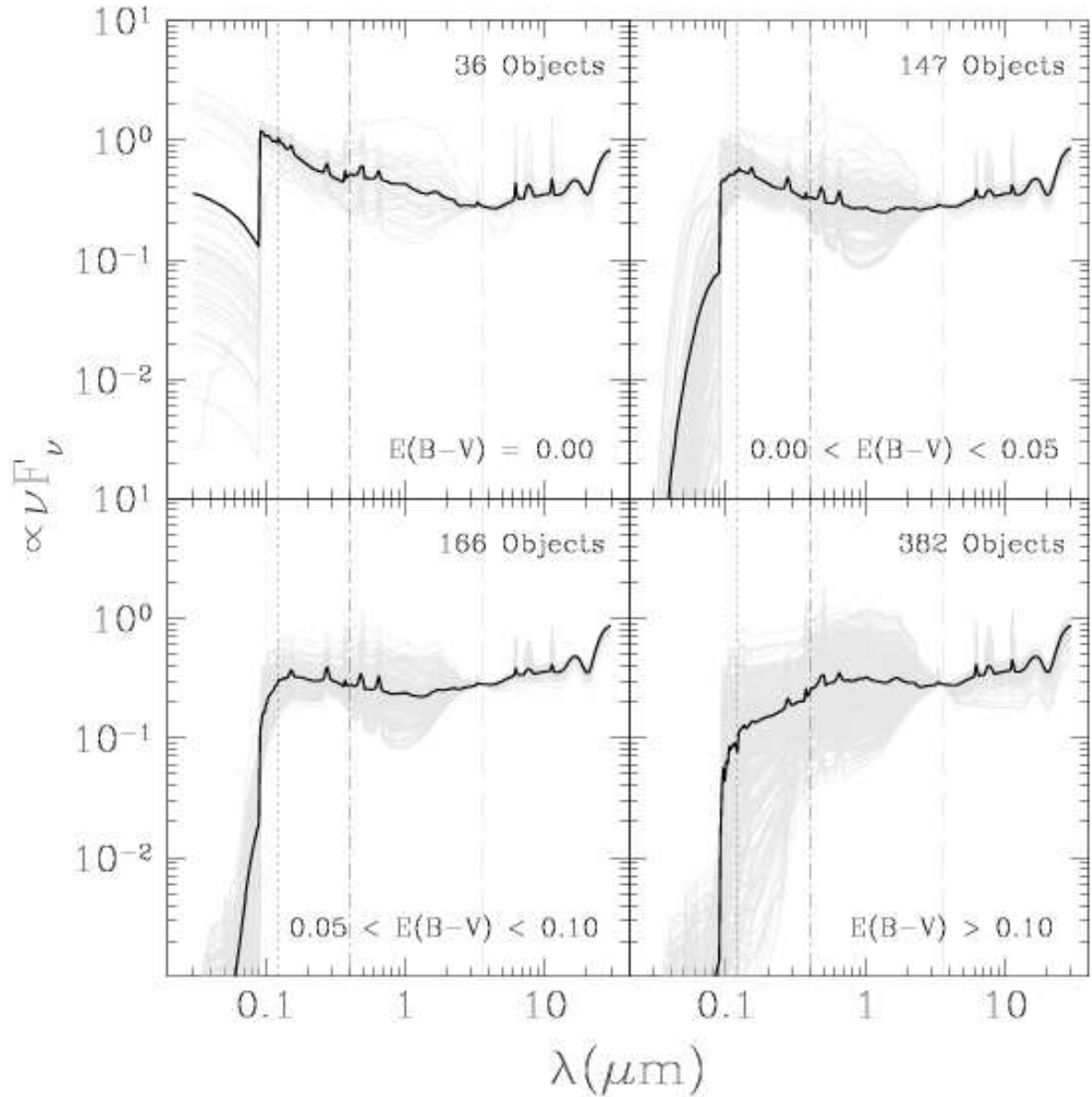}
    \caption{Same as Figure \ref{fg:SEDs_photo_z_good}, but for point
    source AGNs with bad photometric redshifts ($|z_p-z_s|>0.5$). Note
    how these sources have very flat, featureless SEDs compared to
    those in Fig. \ref{fg:SEDs_photo_z_good}. For most of these AGN,
    the Lyman break is not in the range of the data.}
    \label{fg:SEDs_photo_z_bad}    
  \end{center}
\end{figure}

The photometric redshift accuracies for AGNs are much worse. For
extended AGNs, the dispersion is 20\% larger than for galaxies
($\Delta z = 0.050$) and for point source AGNs it is a factor of $\sim
4.5$ worse ($\Delta z = 0.18$). This lower accuracy is similar to the
results of \citet{rowan08} for QSOs in the SWIRE survey. The problem
can be largely attributed to the lack of prominent broad-band features
in the SED of AGNs as compared to galaxies. For many objects, the AGN
continuum combined with that of the host produces an extremely flat
SED, and for these objects the photometric redshift estimate is
dominated by secondary factors like photometry errors and the galaxy
luminosity prior we apply to the hosts. The latter has the effect of
assigning severely underestimated redshift values to many sources,
although removing it improves the median shift but worsens the
accuracy.  Figures \ref{fg:SEDs_photo_z_good} and
\ref{fg:SEDs_photo_z_bad} show the average best fit SEDs for the point
source AGNs with ``good'' and ``bad'' photometric redshifts (defined
by $|z_p - z_s| < 0.2$ and $|z_p - z_s| > 0.5$ respectively) as a
function of the estimated reddening of the AGN. It is very clear that
the objects with good photometric redshifts have significantly
stronger host components than those with bad estimates. In particular,
the objects with good redshifts appear to have a distinctive inflexion
point close to 4000\AA\ which probably constrains the photometric
redshift.

\begin{figure}
  \begin{center}
    \plotone{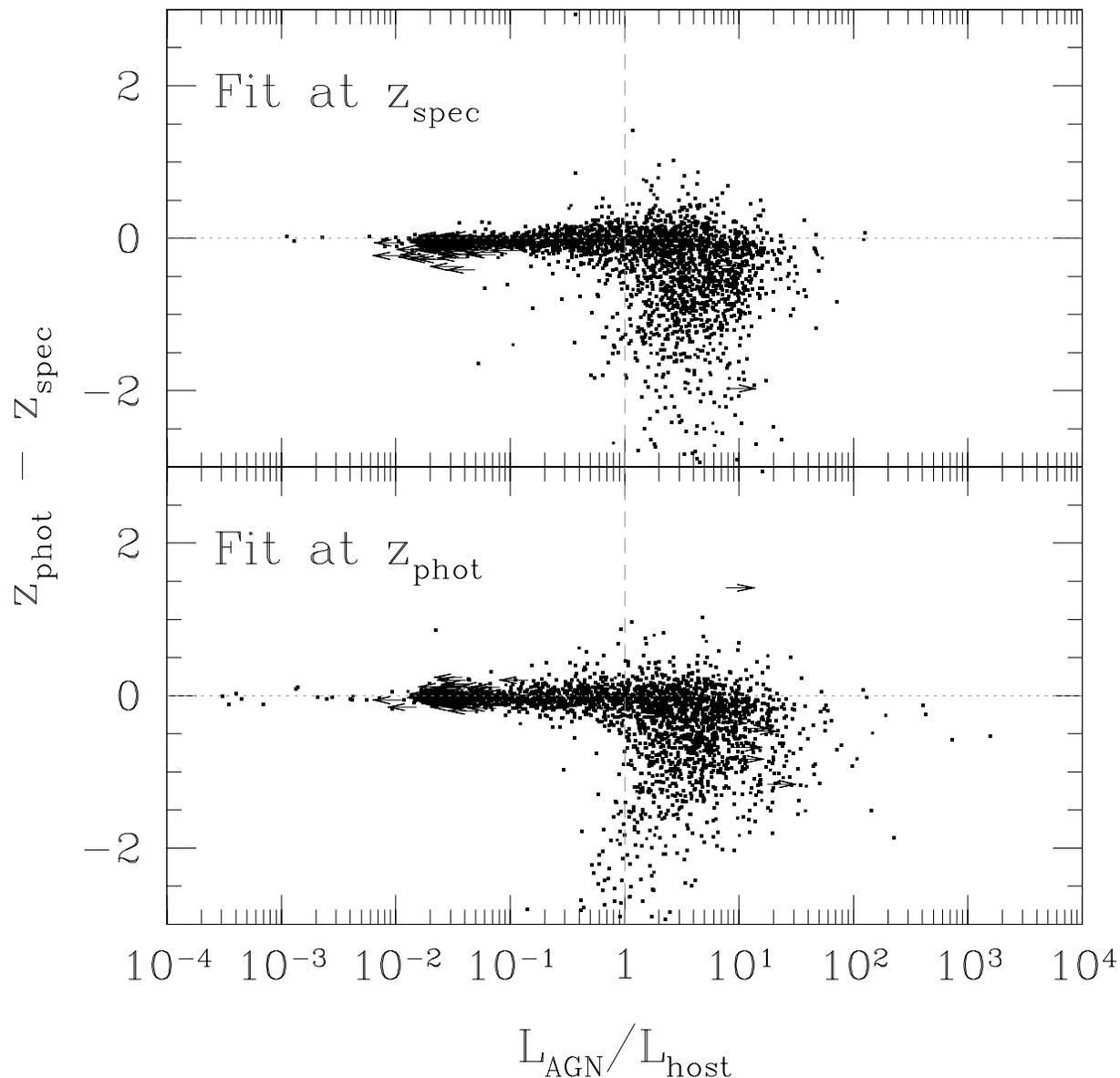}
    \caption{Difference between the photometric and spectroscopic
    redshift of all point source AGNs in our sample as a function of
    the bolometric luminosity ratio between the best fit AGN and host
    components. Arrows show upper or lower limits when either the AGN
    or the host components were not used in the best fit. When the AGN
    is more luminous than its host ({\it{rightward of vertical gray
    dashed line}}), the photometric redshift accuracy is poor.}
    \label{fg:dz_agn_host_ratio}
  \end{center}
\end{figure}

\begin{figure}
  \begin{center}
    \plotone{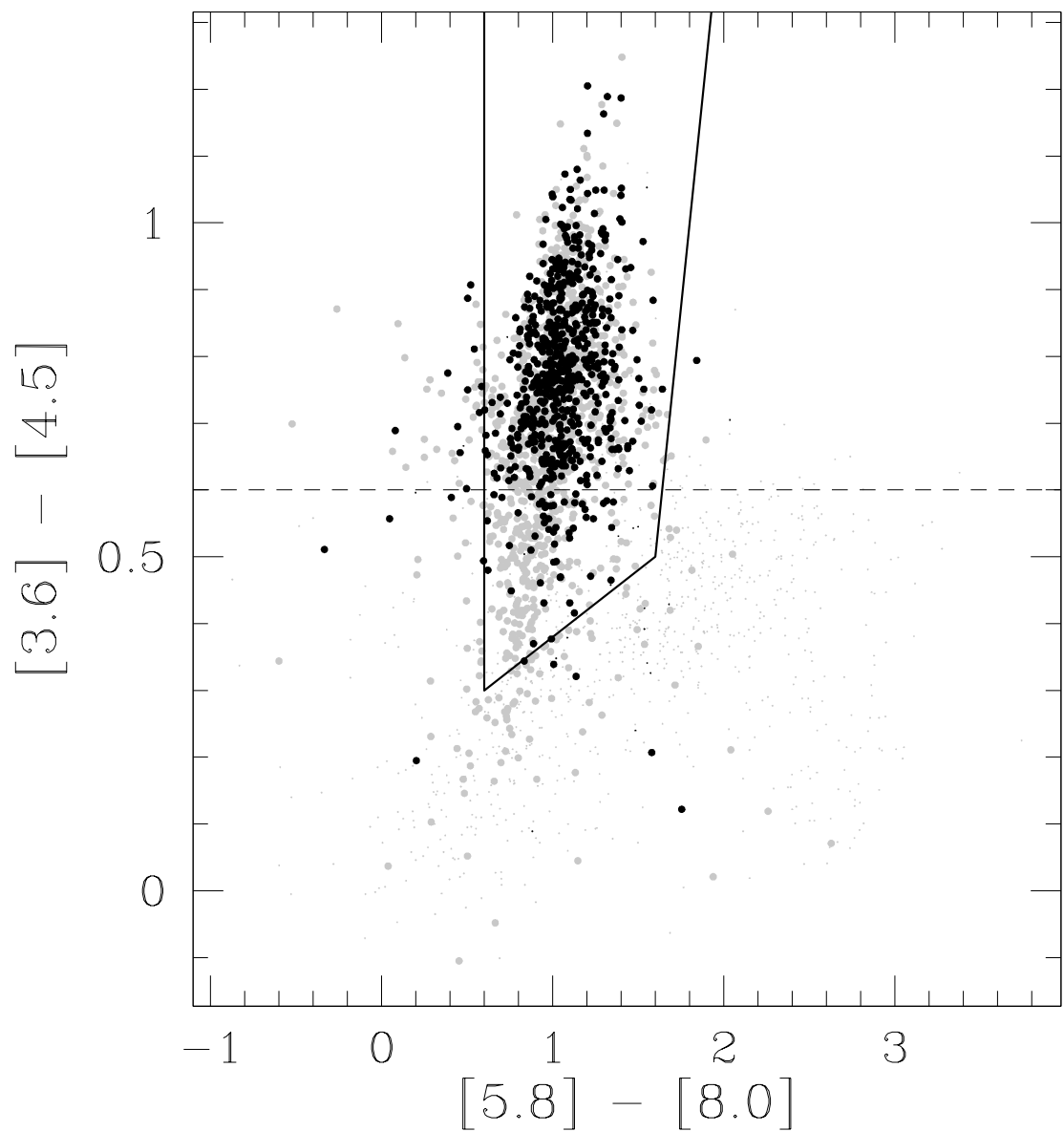}
    \caption{IRAC color-color diagram of all point source AGNs in our
    sample. Black points to the objects with bad photometric redshifts
    ($|z_p-z_s|>0.5$) while the rest are shown in gray. Large points
    mark objects classified by the modified SDSS pipeline to have
    active nuclei, while the rest correspond to the small
    symbols. Notice that the majority of solid circles lie at $\rm
    [3.6] - \rm [4.5]$ colors redder than 0.6. The solid lines mark
    the AGN color selection region of \citet{stern05}.}
    \label{fg:AGN_IRAC_color}
  \end{center}
\end{figure}

To further explore this point, we show in Figure
\ref{fg:dz_agn_host_ratio} the difference between the photometric and
spectroscopic redshifts as a function of the ratio between the
bolometric luminosities of the AGN and host components for all objects
classified as point source AGNs. Note that we have not included the
galaxy luminosity prior when fitting the SED at $z_p$ (although $z_p$
is calculated with the prior), as it is not included for the fits at
$z_s$ to which we are comparing. Independent of whether we estimate
the ratio at $z_s$ or $z_p$, the photometric redshift accuracy is
markedly better when the best fit AGN component is fainter than the
host component. For $L_{\rm AGN}<L_{\rm Host}$, the photometric
redshift dispersion is $\Delta z = 0.041$ for extended AGNs and
$\Delta z = 0.081$ for the point source ones, while when $L_{\rm
AGN}>L_{\rm Host}$, the dispersions rise to 0.145 and 0.224
respectively. Because there is a marked dependence of the photometric
redshift accuracy on the host galaxy strength relative to the AGN SED,
an IRAC color dependence is also expected. Figure
\ref{fg:AGN_IRAC_color} shows the IRAC color-color diagram where the
points have been color-coded according to their photometric redshift
accuracy. Most of the objects with bad redshift estimates lie at
colors redder than [3.6]--[4.5] = 0.6.

\begin{figure}
  \begin{center}
    \plotone{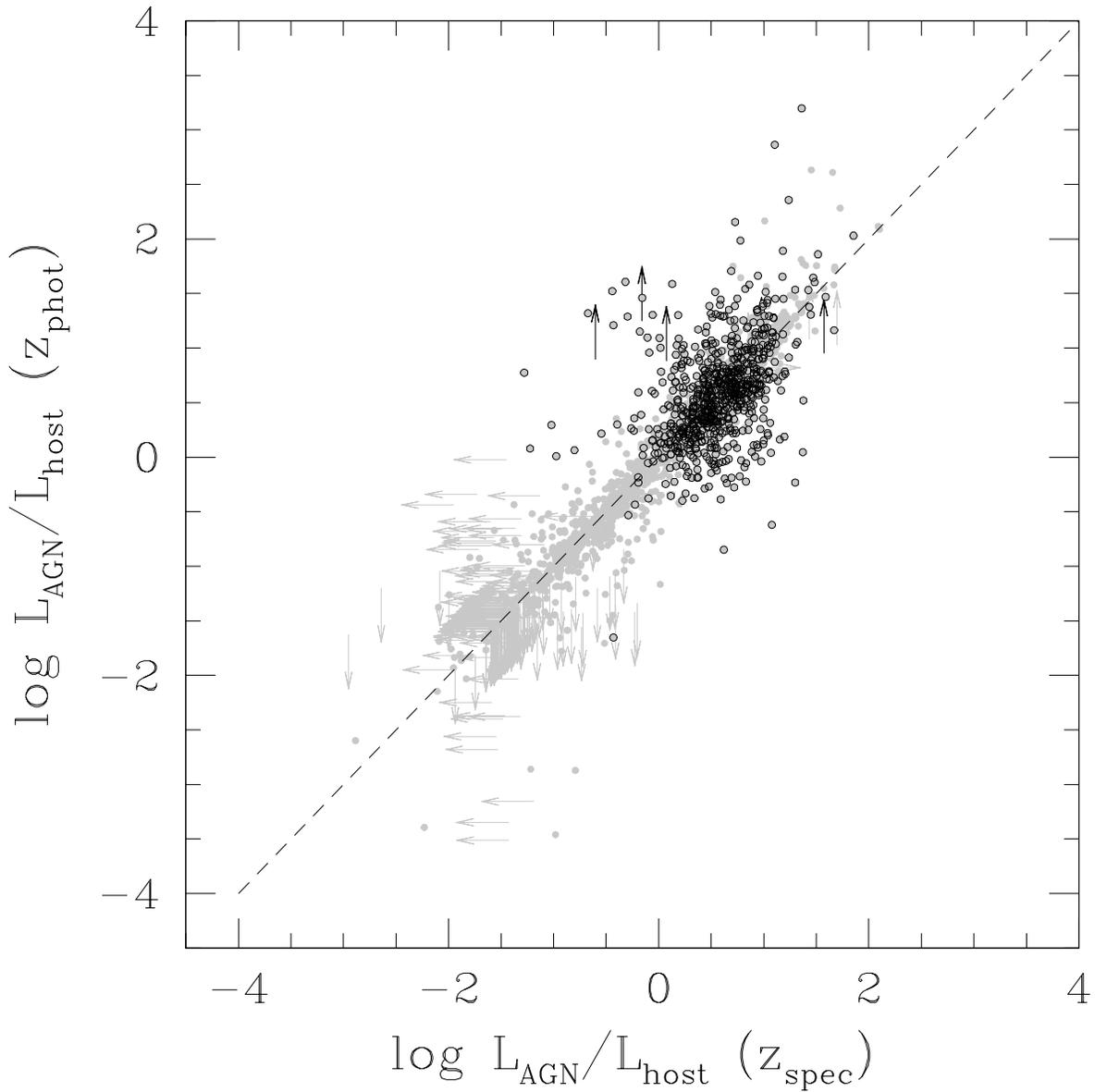}
    \caption{Comparison of the bolometric luminosity ratios between
    the AGN and host components for all point source AGNs ({\it{solid
    gray points}}). Points with black borders correspond to objects
    with bad photometric redshift determinations
    ($|z_p-z_s|>0.5$). When either the AGN or the host component is
    not detected, an arrow shows the upper/lower limit of the
    respective ratio. Note that the ratios are not systematically
    biased when the photometric redshifts are used instead of the
    spectroscopic ones.}
    \label{fg:agn_host_ratio}
  \end{center}
\end{figure}

From Figure \ref{fg:dz_agn_host_ratio} it is possible to see that the
$L_{\rm AGN}/L_{\rm Host}$ ratio is little changed whether we use the
true or photometric redshift of the source. This is more clearly shown
in Figure \ref{fg:agn_host_ratio} and has two interesting
implications. First, since this ratio is not highly dependent on the
photometric redshift, it can be used as a sign of the photo-z's likely
reliability even in the absence of spectroscopic confirmation. And
second, the ability of our templates to measure $L_{\rm AGN}/L_{\rm
Host}$ is not dependent on whether the true redshift of the source is
known, and hence can be used to target (essentially) high Eddington
ratio AGNs using purely photometric redshifts.

At the highest redshifts of our sample ($z>4$), photometric redshifts
are affected by a color degeneracy between the Lyman break at high
redshift and the Balmer break at low $z$, mainly caused by their lack
of detection in the GALEX and Bw bands. It is clear from Figure
\ref{fg:zphot_agns} that if this degeneracy is broken, by, for
example, a combination of deeper UV imaging and stronger upper bounds,
the photometric redshift accuracy for these objects can be very
good. If we recalculate the photometric redshifts for these objects
but only allowing for $z_p > 1$, we obtain an accuracy of
$\sigma_z/(1+z) = 0.072$. Note that this degeneracy is not related to
the fact that we fixed the IGM absorption, as the accuracies for these
high redshifts worsens if we allow the IGM absorption strength to
vary.

We have shown the lack of accuracy in the broad-band
photometric redshifts of AGNs is a product of the lack of spectral
features in their SEDs. Regardless, it is important to understand if
these photometric redshifts really just lack precision or truly lack
accuracy. Based on the probability distribution function (PDF) of each
object, we have calculated 68.3\% and 95.4\% errorbars on the
photometric redshifts. The error bars are asymmetric and have been
defined to include the given percentage of the PDF to either side of
the maximum. For inactive galaxies we find that the average difference
between the 1$\sigma$ (2$\sigma$) upper and lower limits of the
photometric redshifts is 0.08 (0.16), while we find 0.10 (0.21) for
optically extended AGNs and 0.25 (0.54) for optical point source
AGNs. This shows that our photometric redshifts for AGNs lack the
precision of those of inactive galaxies and extended AGNs. If we
recalculate $\sigma_z/(1+z)$ ($\Delta z$) limited to objects where the
68.3\% range of the photometric redshifts is less than 0.1, the range
of the optically extended AGNs, we find 0.062 (0.036) for inactive
galaxies, 0.109 (0.036) for optically extended AGNs and 0.253 (0.118)
for optical point source AGNs. The lack of improvement in these
statistics for the optical point source AGNs suggests that the lack of
spectral features in the SED of the AGNs not only worsens the
precision, but also the accuracy of the photometric redshifts.

The low accuracies of AGN photometric redshifts we have obtained at
$z<4$ can be generally extrapolated to all SED fitting algorithms
relying solely on broad-band data in the UV to mid-IR wavelength
regime, as it is caused by the lack of strong spectral features
inherent to the SEDs of AGNs and not to a method in
particular. Nonetheless, other approaches that do not rely solely on
broad-band SED fits have proven to be significantly more
successful. In particular, the addition of narrow and medium
photometric bands observations can help constrain the photometric
redshifts for AGNs, as shown by \citet{wolf03}, who obtained a
dispersion of $\sigma = 0.03$ for AGNs in the COMBO-17 survey, and by
\citet{salvato09}, who obtained $\sigma/(1+z) \sim 0.015$ for QSOs
with $i^*_{AB} < 24.5$ in the COSMOS field. Narrower bands
significantly enhance the photometric effects of emission
lines. Estimates based on neural networks can also provide accurate
photometric redshifts for AGNs. In particular, \citet{brodwin06} used
this approach to obtain redshifts for QSOs using an earlier version of
the photometry used in this paper for the NDWFS Bo\"otes field. We
have compared their redshift estimates with ours for a sample of 2290
active objects with SEDs dominated by the AGN. While our SED fitting
method gives $\sigma/(1+z) = 0.18$, the neural network approach of
\citet{brodwin06} has an accuracy $\sigma/(1+z) = 0.13$, about 25\%
better.

\subsection{Mid-IR AGN Selection}\label{sec:irac_colors}

Mid-IR photometry has been proven to be a robust and efficient tool to
select AGNs without prior information
\citep[e.g.][]{lacy04,stern05,richards06a}, as their properties at
these wavelengths are typically very different from those of stars and
galaxies. Mid-IR selection methods also avoid some of the biases
against extincted AGN or AGN with $z \sim 3$ found in optical
surveys. In this section we use our galaxy and AGN SED templates to
study the selection of active galaxies by their {\it{Spitzer}} IRAC
colors and possible selection biases as a function of redshift and
reddening of the central engine. In particular, we will focus on the
selection criteria developed by \citet{stern05} and \citet{lacy04}, as
they are the most commonly used AGN selection methods based on IRAC
colors.

\begin{figure}
  \begin{center}
    \plotone{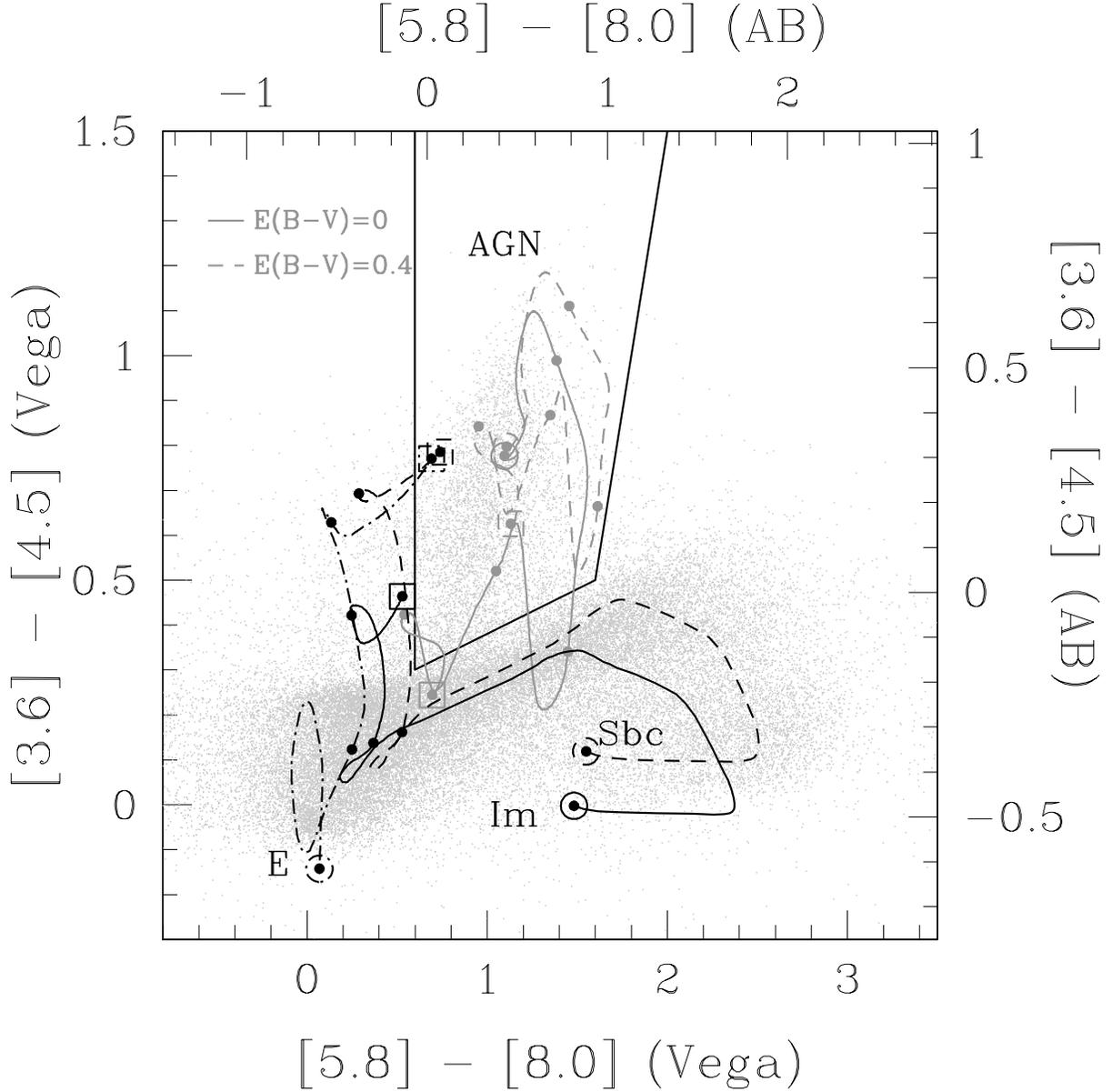}
    \caption{IRAC band color-color diagram of $I\leq 21.5$ SDWFS
    sources ({\it{gray dots}}). Overlaid are the AGN selection region
    of \citet{stern05} ({\it{solid black boundaries}}) and the color
    tracks (as a function of redshift) of our E ({\it{dot-dashed black
    line}}), Sbc ({\it{dashed black line}}) and Im ({\it{solid black
    line}}) galaxy SED templates, and of our AGN template for no
    reddening ({\it{solid gray line}}) and for a reddening of $E(B-V)
    = 0.4$ ({\it{dashed gray line}}). For the galaxy templates, color
    tracks are shown for redshifts between 0--3, while for the AGN
    template they are shown between 0--10. Each heavy dot marks an
    increase of unity in redshift for the galaxies and an increase of
    2 units of redshift for the AGN. For each template, the heavy dot
    surrounded by a circle marks $z=0$ while those surrounded by a
    square mark the terminal redshift ($z=3$ for galaxies and $z=10$
    for the AGN).}
    \label{fg:sdwfs}
  \end{center}
\end{figure}

\begin{figure}
  \begin{center}
    \plotone{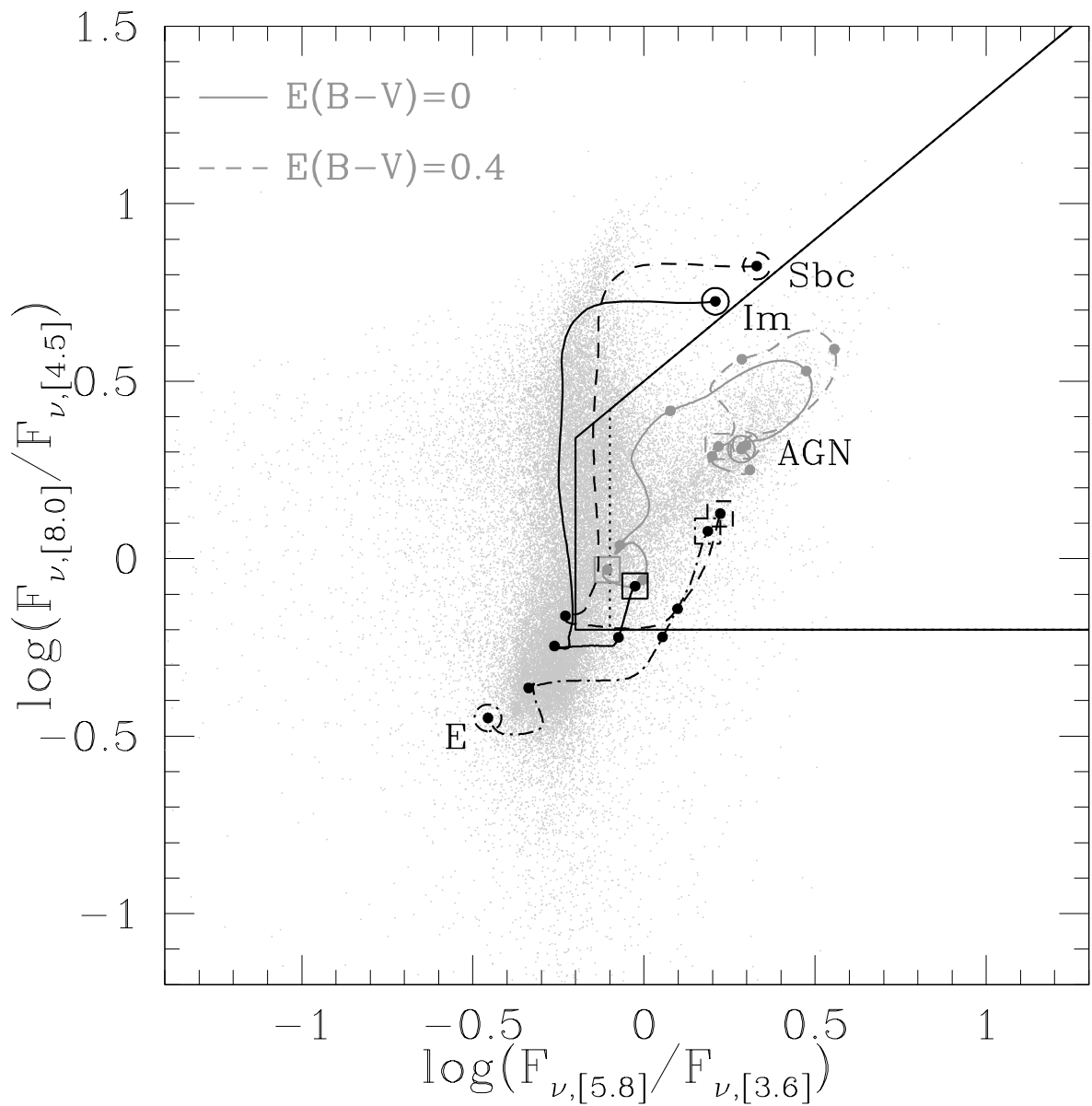}
    \caption{Same as Figure \ref{fg:sdwfs} but in the colors used by
    \citet{lacy04} to define their AGN selection criterion ({\it{solid
    black boundaries}}). We also shown the updated selection criteria
    of \citet{lacy07}, modified to limit the problems with
    contamination by low redshift galaxies ({\it{dotted black
    selection boundaries}}). Line and point styles have the same
    definition as in Figure \ref{fg:sdwfs}.}
    \label{fg:lacy_sdwfs}
  \end{center}
\end{figure}

Figure \ref{fg:sdwfs} shows the mid-IR color space of the SDWFS survey
(only objects with $I<21.5$ to limit the contamination by unflagged
stars) along with the \citet{stern05} quasar selection criterion and
the color tracks of our templates ($z < 3$ for the galaxy templates
and $z<10$ for the AGN). All low redshift galaxies have very similar
[3.6]--[4.5] colors, as this is dominated by their stellar
emission. However, they populate two different regions of the
[5.8]--[8.0] color, where the difference is dust/PAH emission in star
forming galaxies and its absence in ellipticals. As the redshift
increases, the colors converge and follow very similar tracks, as the
bands are redshifted into the predominantly stellar emission at
wavelengths shortward of $\sim 5\mu$m. The IRAC colors of AGNs are, on
the other hand, dominated by hot dust emission at low redshifts/long
wavelengths, and by emission of the accretion disc at higher
redshifts/short wavelengths. As a result, AGNs populate a very
distinct region of the IRAC color-color space from low and
intermediate redshift galaxies, and the \citet{stern05} or
\citet{lacy04} selection criterion approximately delineate this
region. Notice that, as shown in Figures \ref{fg:sdwfs} and
\ref{fg:lacy_sdwfs}, high redshift galaxies ($z\gtrsim 3$) occupy a
very similar region to AGNs in the IRAC colors, and contaminate both
selection criteria. \citet{yun08} has also shown that this is the case
for submillimetre-bright galaxies at $z>2$. However, since AGNs are
commonly much more luminous than galaxies, these issues are only
significant for faint sources.

The \citet{stern05} selection criterion does a remarkably good job of
separating the low and intermediate redshift galaxies from the
AGN. Its only significant problems are the ``blue loop'' out of the
region at $z\simeq 4.5$, where the H$\alpha$ line is redshifted into
the [3.6] band, and the requirement that the AGN, rather than the
host, dominates the mid-IR SED \citep{gorjian08}. A survey with mid-IR
completeness problems near the $z\simeq 4.5$ loop can be easily
handled by including bluer mid-IR objects that also show evidence for
a Lyman limit in the optical. At $z\simeq 7$ our unreddened AGN
template predicts a similar problem when H$\beta$ is redshifted into
the [3.6] band, and the mid-IR colors resemble those $z\sim 1-2$
galaxies. Moderately reddened AGN stay inside the \citet{stern05}
selection boundaries at all redshifts, as illustrated in Figure
\ref{fg:sdwfs} for $E(B-V)=0.4$. With this amount of reddening, when
$H\alpha$ is redshifted into the [3.6] channel, the colors reach the
selection boundary but are always inside of it. Note, however, that
there is a host dependence on the amount of reddening necessary to
stay inside the selection boundaries. As shown by \citet{hickox07}
(see also the top left panel of Fig. \ref{fg:ebv}), more highly
reddened objects can more easily exit the AGN selection boundaries due
to host contamination, as it is harder to meet the requirement that
the AGN dominates the SED. 

The mid-IR AGN selection criterion defined by \citet{lacy04} does not
suffer from the $z\sim 4.5$ incompleteness problem. However, it does
suffer from large amounts of contamination by low redshift star
forming galaxies. Figure \ref{fg:lacy_sdwfs} shows the SDWFS sources
in the color space used by \citet{lacy04} to define their AGN
selection criterion. The AGN colors stay inside the selection
boundaries throughout the whole redshift range from 0--10. However,
star forming galaxies characterized by the Sbc template are also
contained within the selection region at $0.25\lesssim z \lesssim
1$. Note that this also means the requirement that the AGN dominates
the SED is somewhat more relaxed than for the \citet{stern05}
selection criterion. A redder $S_{5.8}/S_{3.6}$ limit would eliminate
most star forming galaxies, but at the cost of introducing the bias
against high redshift QSOs. Hence, the differences between the
\citet{stern05} and \citet{lacy04} AGN selection criteria can be
regarded as a trade-off between completeness and galaxy
contamination. Figure \ref{fg:lacy_sdwfs} also shows an updated
version of this criteria by \citet{lacy07}, on which the vertical
boundary has been shifted by 0.1. This modified boundary keeps the
completeness of \citet{lacy04} and somewhat reduces the galaxy
contamination.

Recently, \citet{richards09} used IRAC observations of SDSS objects to
compare both of the selection criteria, and their results are in very
good agreement with ours. They find that the \citet{stern05} selection
criteria is much less contaminated by inactive galaxies in comparison
to that of \citet{lacy04}, but that the \citet{stern05} selection is
biased against AGNs in the redshift range $3.5 < z < 5$. We find that
this problem is restricted to a somewhat narrower redshift range. In
an earlier work, \citet{donley08} used objects in the GOODS-S field
\citep{goodss} to compare the reliability of these two AGN selection
criteria in addition to the power-law galaxies selection method of
\citet{alonso06} and \citet{donley07}. \citet{donley08} concludes that
the \citet{stern05} selection diagram has more overlap with color
tracks of inactive galaxies than that of \citet{lacy04}, suggesting
that AGN selection by the former is subject to more
contamination. This is possibly due to including significantly fainter
galaxies, but the dominant cause seems to be that their SED models
exaggerate the color space populated by galaxies. As can bee seen in
their Figure 6 for objects with $0.25\leq z\leq 0.75$, few non-AGN
actually lie in the \citet{stern05} region even though the SED tracks
would allow them to do so.

\subsection{Colors of Galaxies and AGNs in WISE}\label{sec:colors}

The Wide-Field Infrared Survey Explorer \citep[WISE;][]{mainzer05}
commenced an all-sky survey in the mid-IR in late 2009. The survey
will be made in 4 bands with effective wavelengths of approximately
3.4, 4.6, 12 and 22$\mu$m, estimated by assuming a source with flux
$F_{\nu} \propto \nu^{-1}$. The two shortest wavelength bands are
similar to the [3.6] and [4.5] channels of IRAC while the longest
wavelength band is similar to the MIPS 24$\mu$m channel. Following the
notation we have adopted for the IRAC bands, we will refer to the WISE
bands by [3.4], [4.6], [12] and [22]. The average survey depth (8
repeated observations, 5$\sigma$) will be 0.12, 0.16, 0.65 and 2.6
mJy, respectively. For comparison, the average depth (6\arcsec\
apertures, 5$\sigma$) of the SDWFS survey in the two shortest
wavelength IRAC bands are 5.2 and 7.2 $\mu$Jy respectively
\citep{ashby09}, about 20 times deeper than the average WISE field.

\begin{figure}
  \begin{center}
    \plotone{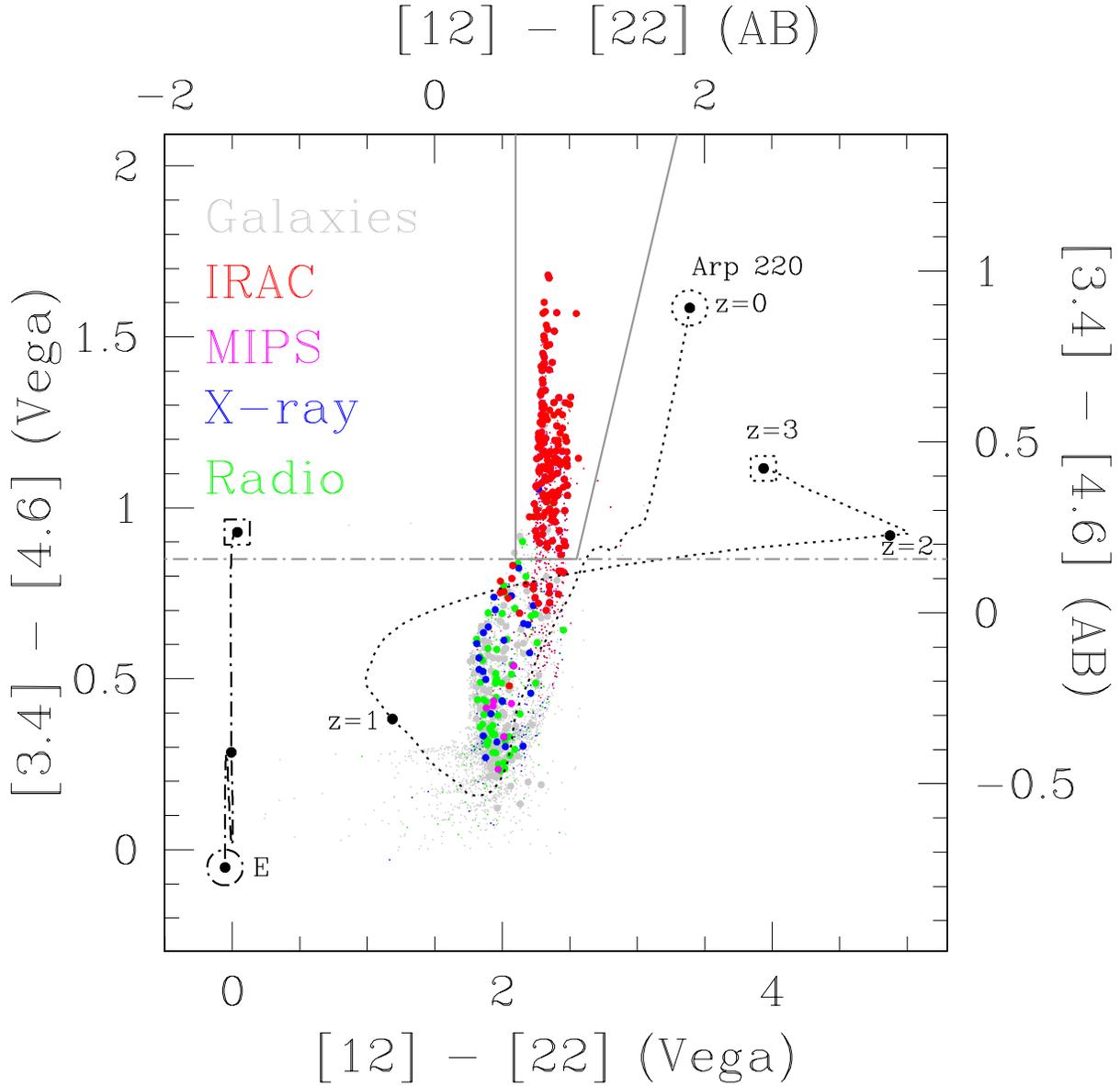}
    \caption{WISE band color-color diagram of all SDWFS sources that
    would be detected in the [3.3] and [4.6] bands ({\it{small
    dots}}), or in all four channels simultaneously ({\it{large
    dots}}). Each color represents objects that were targeted for
    spectroscopy by AGES (but not necessarily observed) as IRAC
    ({\it{red}}), MIPS ({\it{magenta}}), X-ray ({\it{blue}}) and/or
    radio ({\it{green}}) AGNs, or as non-active galaxies
    ({\it{gray}}). These are applied sequentially, so the MIPS points
    are only those that were not IRAC selected and so forth. The
    selection boundaries using all four channels ({\it{solid gray
    line}}) and just the $[3.3]-[4.6]$ color ({\it{dot-dashed gray
    line}}) are also shown. For reference, we show the color tracks of
    the SED of Arp 220 ({\it{dotted black line}}) and of our E
    template ({\it{dot-dashed black line}}) as a function of
    redshift. For each of the SEDs color tracks, heavy dots mark an
    increase in redshift of unity, while the large open circle marks
    $z=0$ and the large open square marks $z=3$ for Arp 220 and $z=2$
    for the E template.}
    \label{fg:wise}
  \end{center}
\end{figure}

\begin{figure}
  \begin{center}
    \plotone{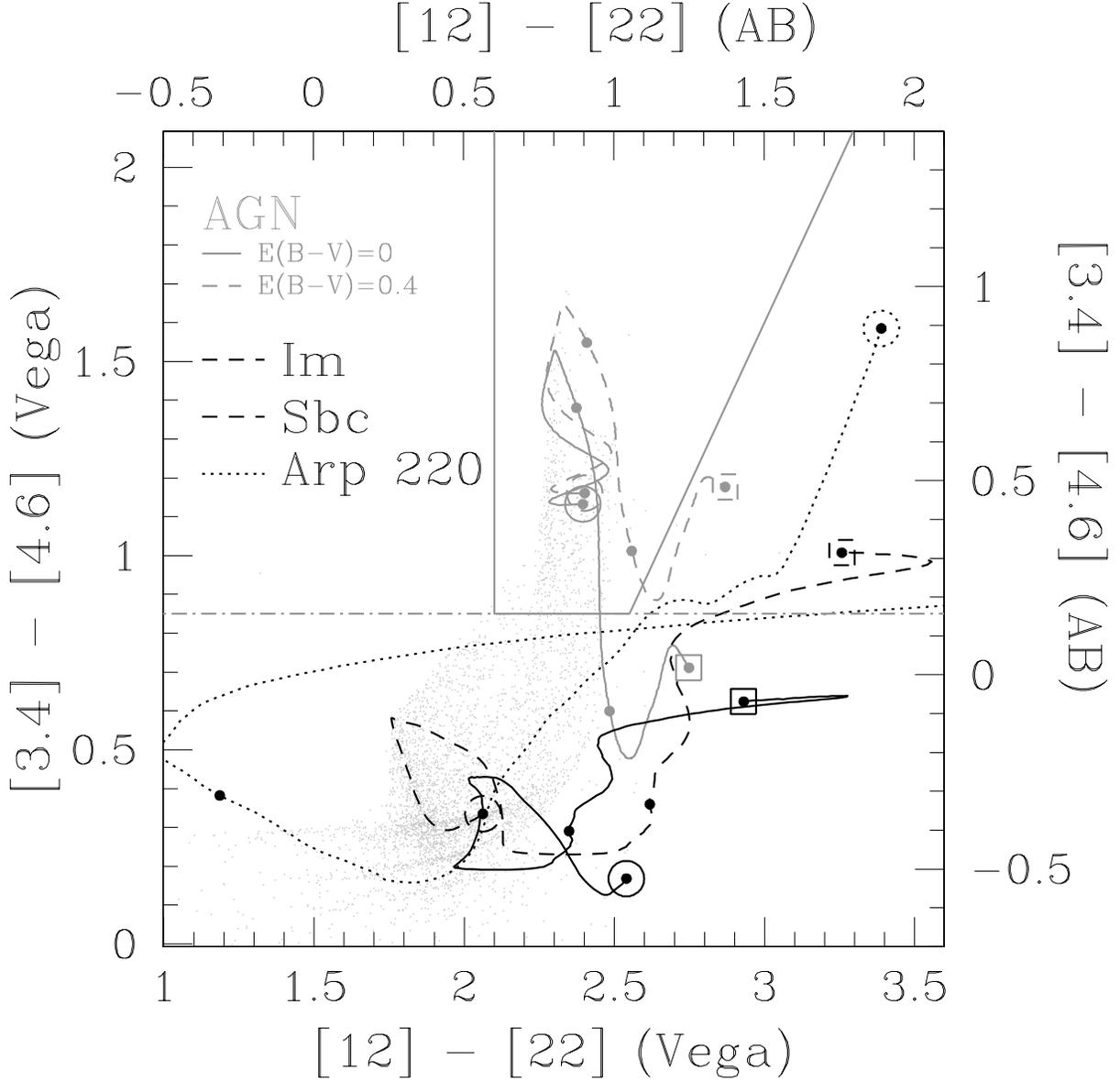}
    \caption{Color tracks of our SED templates and of the SED of Arp
    220, overlaid on top of all SDWFS sources that would be detected
    at [3.3] and [4.6] by WISE ({\it{gray dots}}). The different
    line-styles have the same definition as in Figures \ref{fg:sdwfs}
    and \ref{fg:wise}. Note that, because of the shallow depths of the
    WISE mission, we only show the color tracks up to $z=2$, 6 and 3
    for galaxies AGN and Arp 220 respectively. Most of the Arp 220
    track extends off the red edge of the Figure (see
    Fig. \ref{fg:wise} for the complete color track).}
    \label{fg:wise_tracks}
  \end{center}
\end{figure}

We can combine our object samples and templates to synthesize WISE
observations. For this purpose we use the measured WISE filter curves,
including the detector response (Wright 2009, private
communication). These are shown in Figure \ref{fg:wave_cont}. Table
\ref{tab:kcorr} shows the AB absolute magnitudes of our templates in
the WISE bands as a function of redshift. Figure \ref{fg:wise} shows
the WISE colors as synthesized by our templates for all SDWFS sources
that will be brighter than the WISE [3.4] and [4.6] magnitude limits
and have $I<22.5$. Figure \ref{fg:wise_tracks} shows the color tracks
of our AGN template ($0<z<6$), our star forming galaxy templates
($0<z<2$) and the GRASIL \citep{silva98} Arp 220 SED ($0<z<3)$. The
latter is also shown in Figure \ref{fg:wise} along with the color
tracks of our E template ($0<z<2$). Note that the low redshift
($z\lesssim 0.5$) [12] and [22] fluxes of AGNs are not reliable
because there are too few pure AGNs in our sample at these low
redshifts to stably determine the AGN SED at these wavelengths. The
apparent lack of pure elliptical galaxies in Figure \ref{fg:wise} is
an artifact. Tiny contributions from the star forming SED templates
significantly modify the colors of such galaxies in the two longest
wavelength bands, whether these contributions are real
\citep[e.g. 24$\mu$m excess red galaxies of][]{brand09} or simply an
artifact due to model uncertainties.

Of the 3973 sources in Figures \ref{fg:wise} and \ref{fg:wise_tracks},
3694 have spectroscopic redshifts and we use photo-z's for the
rest. We have eliminated all photo-z objects with fits having
$\chi^2>50$ (63\% of the photo-z sources) in order to limit
contamination by stars. Since the WISE survey is significantly
shallower than SDWFS, essentially all the AGNs at the WISE depths have
spectroscopic redshifts and our results are little affected by any
problems with their photometric redshifts. Objects targeted by AGES as
AGNs are marked in Figure \ref{fg:wise} according to the criterion by
which they were selected. Objects targeted by multiple criteria are
marked only by one criterion, where the priority ordering was IRAC,
MIPS, X-ray and, lastly, radio selection. IRAC selected AGNs clearly
populate a distinct region of the WISE color-color diagram, well
separated from galaxies.

Non-IRAC targeted X-ray and radio candidates have a much larger
scatter and tend to populate a region that extends between the colors
of IRAC selected AGNs and normal star forming galaxies. This is
explained by the requirement that IRAC selected AGN must have higher
Eddington ratios, so that color is dominated by the AGN in the
rest-frame mid-IR. Lower Eddington ratios have colors contaminated by
the host, and hence have mid-IR colors increasingly dominated by stars
and cold dust emission from star formation. X-ray sources are little
affected by host contamination and so are still found when the mid-IR
colors are dominated by the host \citep{gorjian08}, while the WSRT
radio survey is deep enough to detect radio emission associated with
star formation as well as AGN. Note that for $z\gtrsim 0.5$ the host
contamination needed to move Type 1 AGNs out of the selection region
is not just a function of its SED and the Eddington ratio of AGN, but
also of reddening, as both, the [3.4] and [4.6] channels can be
affected by extinction.

In an analogous way to the AGN selection region defined by
\citet{stern05}, we propose the following color-color criterion for
selecting AGNs in WISE observations,
\begin{align}
&[12] -[22]  >  2.1 \label{eq:crit1}\\ 
&[3.4]-[4.6] >  0.85 \label{eq:crit2}\\
&[3.4]-[4.6] >  1.67\  ([12]-[22]) - 3.41 \label{eq:crit3},
\end{align}
\noindent as shown in Figures \ref{fg:wise} and
\ref{fg:wise_tracks}. Like the \citet{stern05} selection method, it
will have little contamination from low and intermediate redshift
galaxies. The left boundary is chosen because there do not seem to be
any AGNs at redder colors either observationally or based on our
templates. The other two boundaries are selected as a compromise
between maximizing the number of WISE selected AGNs and limiting the
contamination by intermediate redshift galaxies and ULIRGS, were we
have characterized the latter by the SED of Arp 220. In particular,
the rightmost boundary separates AGNs from low ($z\lesssim 0.2$) and
high ($z\gtrsim 2$) redshift ULIRGs. ULIRGs are relatively rare and
interesting, so the simple and deeper criterion for selecting AGNs of
\begin{equation}\label{eq:crit4}
[3.4]-[4.6] > 0.85\ ,
\end{equation}
\noindent (the dashed line in Figure \ref{fg:wise}) may be all that is
needed. Note that all WISE selection criteria have significant
problems at $z\gtrsim 3.4$, starting when H$\alpha$ emission is
redshifted into the [3.4] band, analogous to our earlier discussion of
the \citet{stern05} selection criterion. However, in this case,
reddened AGN do not re-enter the selection region at higher
redshifts. WISE data will need to be merged with other information to
separate high redshift QSOs from lower redshift galaxies.

Inside the full selection color region described by equations
(\ref{eq:crit1}), (\ref{eq:crit2}) and (\ref{eq:crit3}), we find 140
Bo\"otes objects flagged as AGN by one of the methods described in
\S\ref{sec:data} and that are bright enough to be detected by WISE in
all of its channels. Since the full NDWFS Bo\"otes field survey area
is approximately 9~deg$^2$, scaling to a full sky survey area we
predict that WISE should find $640,000$ AGNs, corresponding to a
number density of 16~deg$^{-2}$, using our color selection
criteria. If only the criterion of equation (\ref{eq:crit4}) is
applied, we find 375 objects flagged as AGN and bright enough to be
detected in the two shortest wavelength channels of WISE. Again
scaling to the full sky area of WISE, we predict it should find
$1,700,000$ AGNs corresponding to a number density of 42
deg$^{-2}$. \citet{ashby09} estimated a high surface density but used
a bluer cut of [3.4]--[4.6] which will introduce significant
contamination by ULIRGS and star forming galaxies at low and
intermediate redshifts.

\begin{figure}
  \begin{center}
    \plotone{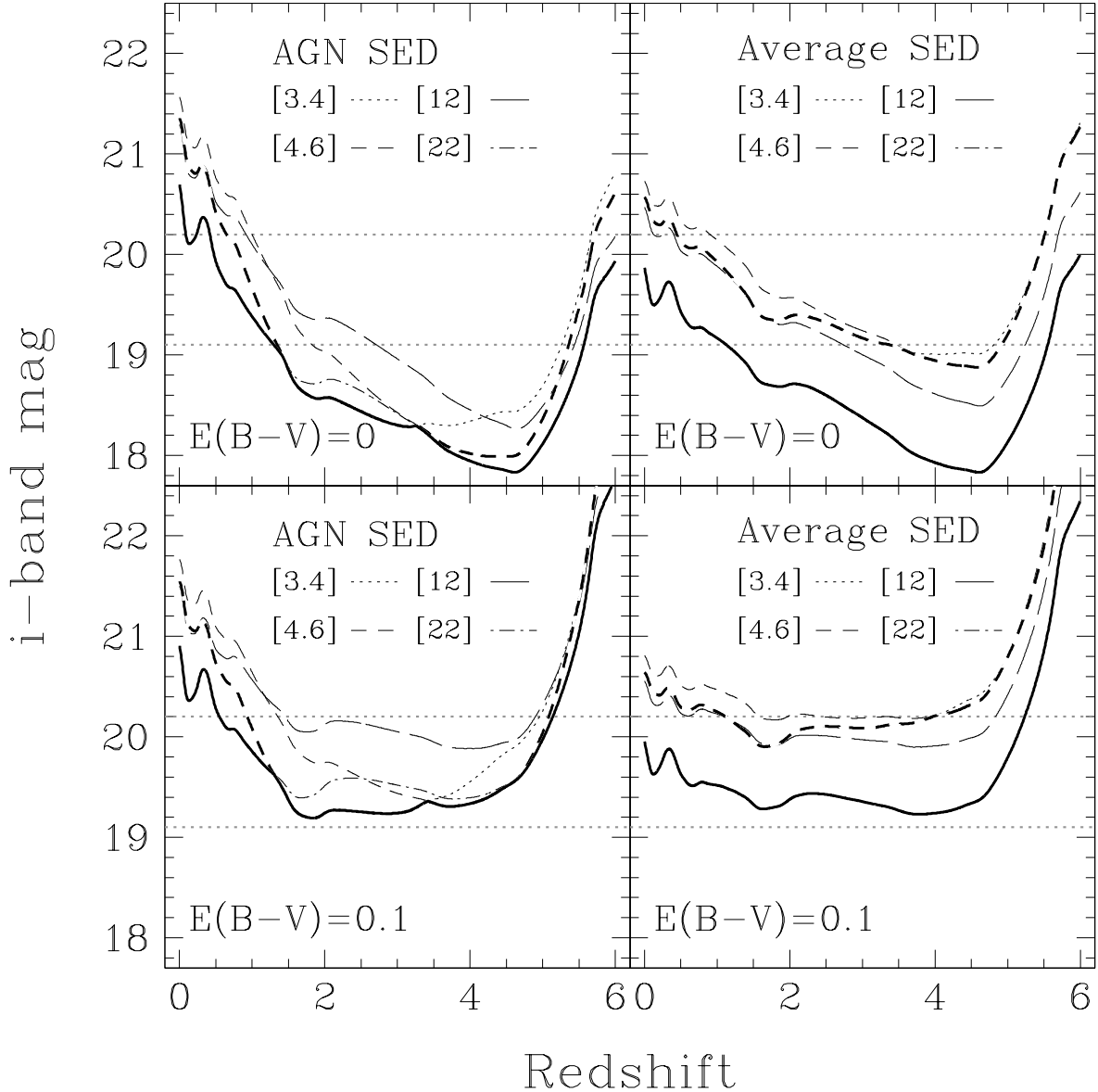}
    \caption{This Figure shows the effective $i$-band magnitude limits
    of the WISE survey for unreddened AGNs ({\it{top}}) and moderately
    reddened AGNs ($E(B-V)=0.1$,{\it{bottom}}), for pure AGN SED on
    the left and the average AGN + Host SEDs of IRAC-selected AGNs on
    the right. Each black line shows the $i$-band magnitude
    corresponding to the depth of each WISE band as a function of
    redshift. The bold solid (bold-dashed) black line shows the
    $i$-band magnitude limit when detection is required in all four
    (two shortest wavelength) WISE bands. The dotted gray lines show
    the SDSS limits for $z<2.3$ ($i < 19.1$) and $z\leq 2.3$
    ($i<20.2$).}
    \label{fg:sdss_wise}
  \end{center}
\end{figure}

These surface densities are higher than those of the SDSS survey, which
finds 11.2~deg$^{-2}$ with $z<2.3$ and $i<19.1$ and 1.03~deg$^{-2}$
and $i<20.2$, but the differences are not a simple change in this
numbers. Figure \ref{fg:sdss_wise} shows the $i$-band magnitude limit
corresponding to the WISE band detection limits for several models of
AGN SEDs. For our pure AGN template without a host or extinction, the
effective $i$-band depth of the WISE sources changes strongly with
redshift, reaching a minimum at $z\sim 4.5$ and then rising again. The
drop at low redshift is due to the v-shaped structure of the SED (see
Fig.~\ref{fg:temps}) - the ratio of optical to mid-IR fluxes rises
with redshift until the minimum at 1$\mu$m enters the mid-IR band
passes at $z\sim 4$, while at the same time the IGM absorption begins
to remove optical flux. Adding a typical level of host emission
markedly reduces these effects by filling in the minimum of the SED,
to make the effective optical magnitude limit of WISE for the
[3.4]/[4.6] bands deeper than $i=19.1$ at almost all redshifts. Adding
a small amount of extinction to suppress the optical relative to the
mid-IR has similar effects. Thus, the WISE sample will complement the
SDSS samples.

\section{Conclusions}

We have created an optimized basis of low resolution SED templates for
AGNs and galaxies in the wavelength range from 0.03--30$\mu$m derived
from the extensive multi-wavelength observations of the NDWFS Bo\"otes
field and the AGES spectroscopic survey. This basis of templates
consists of three galaxy SED templates that resemble elliptical,
spiral and irregular galaxies, and an AGN SED template with variable
reddening. Our model also includes IGM absorption of variable strength
for high redshift sources. The templates and source codes needed to
synthesize other bands and determine photo-zs are publicly
available\footnote{Templates and codes available at
\url{www.astronomy.ohio-state.edu/~rjassef/lrt}}.

In this paper we have investigated three applications of our
templates, namely the estimation of photometric redshifts for galaxies
and AGNs, the reliability of AGN selection in the IRAC bands, and the
prediction of the colors of these objects in the WISE survey as a
function of redshift. In a subsequent paper \citepalias{assef09} we
have used them to study the luminosity function of mid-IR selected
quasars from $z=0 - 5.85$.

The photometric redshift accuracy of our templates for galaxies is
good and comparable to other methods in the literature and to those in
\citetalias{assef08}. For AGNs, however, the accuracy is much lower,
in particular for those that have little contribution from the host
galaxy. This is in agreement with the results of \citet{rowan08} for
QSOs in the SWIRE survey. The photometric redshift accuracy is highly
correlated with the relative luminosities of the AGN and host
components ($L_{\rm AGN}/L_{\rm Host}$), in the sense that the
accuracy is lower for higher AGN fractions. In this limit, the SEDs of
AGNs are flat, leaving no strong features for determining the photo-zs
in broad-band photometry.  When only objects where $L_{\rm AGN}<L_{\rm
Host}$ are considered, the photometric redshift accuracy is comparable
to that of normal galaxies. We find that the estimate of $L_{\rm
AGN}/L_{\rm Host}$ is robust even when the photo-z is inaccurate.

We have also used our templates to study the color distribution of
galaxies and AGNs in the IRAC bands as a function of redshift and AGN
reddening, and how this affects the selection criteria of
\citet{lacy04} and \citet{stern05}. In particular, we have shown that
the \citet{stern05} criterion suffers from significant incompleteness
at $z\simeq 4.5$, due to the broad H$\alpha$ line being redshifted
into the [3.6] band, but that it is little contaminated by low and
intermediate redshift galaxies. We have also shown that the selection
criterion of \citet{lacy04} is less subject to these sources of
incompleteness but it is heavily contaminated by low redshift
star-forming galaxies. Moreover, shifting the boundaries of the
\citet{lacy04} selection region to limit the contamination by galaxies
creates similar incompleteness issues to those of the \citet{stern05}
criterion. The differences between the two selection methods can be
regarded as a trade-off between completeness and galaxy contamination.

We defined two simple AGN selection criteria for the WISE satellite
mission by synthesizing the WISE bands for all sources in the much
deeper SDWFS survey of the Bo\"otes field using our template
models. The absolute magnitudes of our templates in the WISE bands as
a function of redshift are provided in Table \ref{tab:kcorr}. A
restrictive criteria using all 4 WISE bands would identify 140 AGN in
the Bo\"otes field, while one based only on the 2 shorter wavelength
bands would identify 375. The main differences are that the more
restrictive 4 filter criterion would avoid contamination by ULIRGs but
is shallower because it requires detection in the less sensitive
longer wavelength bands. Extrapolating from the $\sim 9~\rm deg^2$
Bo\"otes field to the full sky, the two criteria would identify
$640,000$ and $1,700,000$ AGNs respectively but largely with
$z<3.4$. Additional information is needed to identify objects with
$z>3.4$. These criteria should work also reasonably well in regions of
high stellar density but little star formation \citep[see e.g.][on applying
the \citealt{stern05} approach to find QSOs behind the Magellanic
Clouds]{kozlowski09}.

\acknowledgments 

We would like to thank all the people in the NDWFS, FLAMEX and SDWFS
collaborations that did not directly participate in this work. We
would also like to thank Edward L. Wright for providing us the filer
functions of the WISE mission passbands. We thank the anonymous
referee for providing useful comments and suggestions that improved
this paper. Support for MB was provided by the W. M. Keck
Foundation. The work of DS was carried out at Jet Propulsion
Laboratory, California Institute of Technology, under a contract with
NASA. The AGES observations were obtained at the MMT Observatory, a
joint facility of the Smithsonian Institution and the University of
Arizona. This work made use of images and/or data products provided by
the NOAO Deep Wide-Field Survey \citep{ndwfs99,jannuzi05,dey05}, which
is supported by the National Optical Astronomy Observatory
(NOAO). This research draws upon data provided by Dr. Buell Jannuzi
and Dr. Arjun Dey as distributed by the NOAO Science Archive. NOAO is
operated by AURA, Inc., under a cooperative agreement with the
National Science Foundation.

\begin{deluxetable}{c c c c c c}

\tablehead{$\lambda$($\mu$m) & \multicolumn{5}{c}{$F_{\nu}$ (erg/s/cm$^2$/Hz)}\\  & AGN ($\times 10^{-11}$) & AGN 2 ($\times 10^{-14}$) & E ($\times 10^{-16}$) & Sbc ($\times 10^{-17}$) & Im ($\times 10^{-15}$)}

\tablecaption{SED Templates \label{tab:spectab}}
\tabletypesize{\small}
\tablewidth{0pt}
\tablecolumns{6}

\startdata

    0.0303 &     0.9662 &     5.5387 &     0.9659 &     1.9502 &     1.1327 \\
    0.0311 &     0.9838 &     5.7997 &     1.0250 &     2.1322 &     1.2553 \\
    0.0318 &     1.0008 &     6.1105 &     1.0829 &     2.2252 &     1.3064 \\
    0.0325 &     1.0153 &     6.4724 &     1.1465 &     2.3982 &     1.4211 \\
    0.0333 &     1.0299 &     6.8141 &     1.2075 &     2.5162 &     1.4919 \\

\enddata

\tablecomments{Electronic table that presents the flux per unit
frequency $F_{\nu}$ of our templates. The column marked as ``AGN''
contains our standard AGN template while that marked ``AGN 2''
contains the SED derived by using the \citet{richards06a} SED as a
starting point. Templates are normalized to be at a distance of 10~pc
and to have an integrated luminosity between the wavelength boundaries
of $10^{10} L_{\odot}$.}
\end{deluxetable}

\begin{deluxetable}{l c c c c c c c c c c c c c c c c c c c c c c c}

\tablecaption{SED Templates Absolute Magnitudes\label{tab:kcorr}}

\rotate
\setlength{\tabcolsep}{0.03in} 
\tablehead{$z$ & Template & NUV & FUV & $B_W$ & $B$ & $V$ & $R$ & $I$ & $z$ & $J$ & $H$ & $Ks$ & $K$ & [3.6] & [4.5] &
[5.8] & [8.0] & 24$\mu$m & [3.4] & [4.6] & [12] & [22] & DM}
\tabletypesize{\scriptsize}
\tablewidth{0pt}
\tablecolumns{24}

\startdata

0.0 &     1&  --15.78&  --15.69&  --15.48&  --15.36&  --15.34&  --15.69&  --15.73&  --15.32&  --16.65&  --17.86&  --19.01&  --19.10&  --20.83&  --21.62&  --22.49&  --23.53&  --27.67&  --17.96&  --18.40&  --19.70&  --20.60&  \nodata\\
0.0 &     2&  --12.00&  --13.63&  --18.51&  --18.59&  --19.50&  --20.13&  --20.78&  --20.52&  --21.75&  --22.50&  --22.67&  --22.65&  --22.77&  --22.64&  --22.57&  --22.59&  --22.50&  --20.06&  --19.32&  --17.42&  --15.87&  \nodata\\
0.0 &     3&  --14.66&  --15.86&  --18.26&  --18.28&  --19.00&  --19.59&  --20.20&  --19.94&  --21.22&  --22.02&  --22.23&  --22.22&  --22.80&  --22.93&  --24.57&  --26.06&  --29.56&  --20.04&  --19.69&  --22.04&  --22.60&  \nodata\\
0.0 &     4&  --17.56&  --18.05&  --19.24&  --19.22&  --19.57&  --19.88&  --20.22&  --19.82&  --20.65&  --21.12&  --21.28&  --21.27&  --21.67&  --21.68&  --23.13&  --24.57&  --28.31&  --18.95&  --18.42&  --20.43&  --21.47&  \nodata\\
0.1 &     1&  --15.87&  --15.76&  --15.63&  --15.50&  --15.62&  --15.72&  --15.91&  --15.38&  --16.59&  --17.70&  --18.89&  --18.98&  --20.80&  --21.59&  --22.44&  --23.52&  --27.39&  --17.92&  --18.37&  --19.64&  --20.42&   38.23\\
0.1 &     2&  --12.02&  --13.16&  --17.93&  --18.08&  --19.36&  --20.02&  --20.70&  --20.44&  --21.80&  --22.49&  --22.86&  --22.89&  --23.01&  --23.02&  --22.86&  --22.89&  --22.82&  --20.28&  --19.69&  --17.74&  --16.19&   38.23\\
0.1 &     3&  --14.65&  --15.66&  --17.95&  --18.01&  --18.88&  --19.50&  --20.15&  --19.87&  --21.23&  --22.01&  --22.45&  --22.46&  --22.98&  --23.11&  --23.72&  --26.16&  --29.41&  --20.20&  --19.82&  --22.09&  --22.48&   38.23\\
0.1 &     4&  --17.61&  --18.04&  --19.10&  --19.10&  --19.66&  --19.91&  --20.28&  --19.87&  --20.80&  --21.24&  --21.50&  --21.52&  --21.91&  --21.91&  --22.37&  --24.69&  --28.29&  --19.15&  --18.61&  --20.47&  --21.38&   38.23\\
0.2 &     1&  --16.00&  --15.88&  --15.81&  --15.67&  --15.69&  --15.85&  --16.18&  --15.49&  --16.55&  --17.58&  --18.76&  --18.86&  --20.75&  --21.58&  --22.39&  --23.52&  --27.30&  --17.87&  --18.35&  --19.61&  --20.49&   39.87\\

\enddata

\tablecomments{The electronic table supplies the absolute magnitude of
each SED template (AGN: 1, E: 2, Sbc: 3 and Im: 4) for a broad range
of photometric bands as a function of redshift, along with the
distance modulus DM. A complete version of this table can be found in
the electronic edition of the journal. Templates have been normalized
to a ``bolometric'' luminosity of $10^{10}L_{\odot}$. The absolute
magnitude we present here corresponds to the canonical definition of
the absolute magnitude \citep[as in, for example, eqn. 26
of][]{hogg99} plus the $K$ correction term. This allows the
calculation of photometric redshifts and $K$ corrections from the
table. This is equivalent to Tables 3 and 4 of \citetalias{assef08}
(see their caption for details). All magnitudes are in the VEGA system
except for NUV, FUV, $z$ and the WISE bands, which are in the AB
system. Note that no template incorporates IGM absorption and the AGN
template is unreddened. For calculating photometric redshifts and
$K$-corrections allowing for variable IGM absorption and obscuration
of the AGN, public numerical routines are provided on-line at
{\tt{http://www.astronomy.ohio-state.edu/$\sim$rjassef/lrt}}.}

\end{deluxetable}

\begin{deluxetable}{l c c c c c c}

\tablehead{Sample & $\sigma_z/(1+z)$ & $\Delta z$ & 68.3\% & 95.5\% &
99.7\% & Median}

\tablecaption{Photometric Redshift Estimation Summary.\label{tab:photoz}}
\tabletypesize{\small}
\tablewidth{0pt}
\tablecolumns{7}

\startdata

All               & 0.155 & 0.057 & 0.049 & 0.240 & 0.721 & -0.020 \\
Point Source AGNs & 0.261 & 0.184 & 0.174 & 0.544 & 0.777 & -0.121 \\
Extended AGNs     & 0.149 & 0.050 & 0.049 & 0.175 & 0.812 & -0.006 \\  
Galaxies          & 0.128 & 0.041 & 0.042 & 0.127 & 0.429 & -0.017 \\

\enddata

\tablecomments{Summary of the photometric redshift calculations for
our main catalog and each sub-sample discussed in
\S\ref{ssec:res_zphot}. The table shows for each case the value of
$\sigma_z/(1+z)$ (as defined by eqn. [\ref{eq:sigz}]), $\Delta z$ (the
95\% clipped distribution $\sigma_z/(1+z)$), the ranges of
$|z_p-z_s|/(1+z)$ encompassing 68.3\%, 95.5\% and 99.7\% of the
distribution and the median value of $z_p-z_s$.}

\end{deluxetable}

\clearpage

\clearpage

\end{document}